\documentclass[10pt,aps,prd,twocolumn,notitlepage,nofootinbib,floatfix,
superscriptaddress]{revtex4-1}

\usepackage{amsmath, amsthm, amssymb, amsfonts, amsbsy, mathrsfs}
\usepackage{resizegather}
\usepackage{xcolor}
\usepackage{calligra,bm}
\usepackage{stmaryrd}

\usepackage{graphicx}
\usepackage[hidelinks,bookmarks=true]{hyperref} 
\hypersetup{pdfstartview=FitH,pdfhighlight=/O,colorlinks=true}

\begin{document}

\title{Shielding a charged black hole}

\author{Justin C. Feng}
\email{justin.feng@tecnico.ulisboa.pt}
\affiliation{CENTRA, Departamento de F\'{\i}sica, Instituto Superior
T\'ecnico -- IST, Universidade de Lisboa -- UL, Avenida Rovisco Pais 1,
1049 Lisboa, Portugal}

\author{Sumanta Chakraborty}
\email{tpsc@iacs.res.in}
\affiliation{School of Physical Sciences, Indian Association for the 
Cultivation of Science, Kolkata-700032, India}

\author{Vitor Cardoso}
\email{vitor.cardoso@tecnico.ulisboa.pt}
\affiliation{Niels Bohr International Academy, Niels Bohr Institute, 
Blegdamsvej 17, 2100 Copenhagen, Denmark}
\affiliation{CENTRA, Departamento de F\'{\i}sica, Instituto Superior 
T\'ecnico -- IST, Universidade de Lisboa -- UL,
Avenida Rovisco Pais 1, 1049 Lisboa, Portugal}

%\preprint{}

%\date{\today}

%=======================================================================

\begin{abstract}
We describe a shielding mechanism for a charged black hole immersed in a
background involving charged matter fields, solely arising from the
Einstein-Maxwell field equations. In particular, we consider a charged
generalization of the Einstein cluster, that is a charged black hole
surrounded by an effective fluid model for a partially charged dust
cloud. We show that the shielding mechanism, arising thereof, is generic
and appears in a different parametrization of the problem as well. In
this process, we provide the most general electrovacuum solution in a
spacetime region devoid of charges, but in the presence of a static and
spherically symmetric charge distribution elsewhere. Side by side, we
also introduce a convenient parametrization, providing the global
solution of the Einstein-Maxwell's field equations in the presence of a
charged black hole within the environment of charged fluid. We also
comment on the nature of the photon sphere, shadow radius and the
eikonal quasinormal modes in the Einstein-Maxwell cluster. 
\end{abstract}

%=======================================================================

\maketitle

%=======================================================================

%-----------------------------------------------------------------------
%-----------------------------------
%-----------------
%--------
%---
%-
%
%
%-
%---
%--------
%-----------------
%-----------------------------------
%-----------------------------------------------------------------------

%=======================================================================
\section{Introduction}
%=======================================================================
Black holes (BHs) are among the most fascinating objects in modern
physics, connecting the classical and quantum domains of gravity
\cite{Hawking:1975vcx,Mathur:2009hf,Chakraborty:2017pmn}. Though these
are the simplest solutions of the gravitational field equations, they
hide singularities, where the classical laws of Physics break down
\cite{Ashtekar:2022oyq,Christodoulou:1991yfa,Hawking:1976ra,Penrose:1964wq}.
The most remarkable property is that classical BHs inherit a special
surface, known as the event horizon, acting as a one-way membrane.
Experimental verification of the existence of such an event horizon
would also provide conclusive evidence for BHs in nature
\cite{Visser:2008rtf,Giddings:2017jts,Cardoso:2016rao,Cardoso:2017cqb}
(for alternatives to the BH paradigm, see
\cite{Liebling:2012fv,Mazur:2004fk,Mathur:2005zp,Seidel:1991zh,Damour:2007ap,Barcelo:2009tpa,Maselli:2017cmm,Brito:2015yfh,Holdom:2016nek,Abedi:2016hgu,Chakraborty:2022zlq}).

Recent years have seen significant progress in these directions. The
event horizon telescope has already probed the location of the photon
sphere, also known as the unstable circular photon orbits, through the BH shadow
measurements of the central supermassive compact objects in the galaxies
M87 and our own Milky Way
\cite{EventHorizonTelescope:2019dse,GRAVITY:2020gka}. The gravitational
wave detectors as well have probed the photon sphere
\cite{Cardoso:2008bp}, through the lowest lying quasinormal modes
\cite{Berti:2009kk} of the perturbed BH, arising from the merger of
binary BHs  \cite{LIGOScientific:2016aoc,LIGOScientific:2020ibl}. It is
expected that future gravitational wave detectors will be able to probe
various features of these perturbations, including nonlinearities and
other multipoles to reveal finer details, thus settling the question of
whether these compact objects are indeed BHs or not
\cite{Cardoso:2016rao,Cardoso:2016oxy,Cardoso:2017cqb,Barack:2018yly,Cardoso:2019rvt,Cotesta:2022pci,Cheung:2022rbm,Mitman:2022qdl}.

The simplicity of BHs is due to the fact that these objects in general
relativity are described by three parameters alone---the mass, the
angular momentum, and the electric charge. In theories beyond general
relativity, there can be additional parameters and searching for these
additional hairs is one of the prime aims of any astrophysical tests of
gravity, including the gravitational wave observations
\cite{Berti:2015itd,Babichev:2016rlq,Yunes:2011we,Pani:2011xm,Bhattacharya:2016naa,Barack:2018yly}.
However, within the purview of general relativity one always works with
the Kerr metric for describing any astrophysical BHs, i.e.,
astrophysical BHs are assumed to have only mass and angular momentum,
but zero electric charge. Even though pair-production or Hawking
radiation are able to discharge BHs to a good extent
\cite{Cardoso:2016olt}, there does not seem to be any appropriate
mechanism, which can make the use of electric charge completely
irrelevant in astrophysical contexts from first principles, i.e.,
starting from the Einstein-Maxwell field equations.\footnote{Perhaps it
is worth mentioning here some recent work on mergers of charged black
holes  \cite{Pina:2022dye,Luna:2022udb,Bozzola:2021elc,Bozzola:2020mjx},
which may provide one possible mechanism.} This is the gap we wish to
fill in this work.

For that purpose, we start with a charged BH surrounded by an
anisotropic charged fluid, in a static and spherically symmetric
configuration. This is typical of any galactic system, where a central
massive BH would be surrounded by a plasma, having some distinct mass
profiles \cite{Hernquist:1990be,Baes:2004pg}. If the central BH is
charged, the plasma is expected to exhibit overdensity for one species
of charged particle, and hence acquire an overall charge. This is also
expected from the fact that the conductivity of the plasma is supposed
to mitigate the electric field from the central BH. The effect of the
matter distributions surrounding Schwarzschild and charged BHs has
already been studied in
\cite{Einstein:1939ms,Cardoso:2021wlq,10.1143/PTP.65.1281}, where an
exact solution of the Einstein equations describing the surrounding
matter (an ``Einstein cluster''), as well as the central BH, have been
obtained. In this article, we provide a generalization of the solution
considered in \cite{Cardoso:2021wlq} for the case of a charged BH, but
also demonstrate the screening mechanism at work, i.e., why isolated
charged BHs are not relevant in astrophysical contexts. In particular,
we will show that a charged BH surrounded by a spherically symmetric
charged cloud generically results in a Schwarzschild BH, seen from a
large distance. 

The paper is organized as follows: In Sec. \ref{Basic_Eqs}, we first
present the general structure of the gravitational and electromagnetic
field equations describing a charged BH surrounded by charged
anisotropic matter distribution and then specialize to the case of a
static and spherically symmetric situation in Sec.
\ref{Sec:Basic_Eqs_Sph}. Subsequently, we present the screening
mechanism for a given profile of the number density of the distribution
of charged matter in Sec. \ref{Sec:Debye} where the screening of the
BH charge can be seen explicitly. Then, we propose a parametrization
for the charged BH surrounded by the charged matter distribution in
Sec. \ref{Sec:BH_param}, which also shows a screening behavior, as
described in Sec. \ref{Sec:PhysProps}. Then we finish with our
conclusion and some physical properties of the solution derived above.
\\
\\
\emph{Notation and Conventions:} We use units where the fundamental
constants $G$ and $c$ have been set to unity, i.e., $G=1=c$. We also use
geometrized units in the electromagnetic sector, e.g., we choose the
magnetic permeability in vacuum to be $\mu_0=4\pi$ and then the
electric permittivity in vacuum becomes $\epsilon_{0}=(1/4\pi)$.

%=======================================================================

%-----------------------------------------------------------------------
%-----------------------------------
%-----------------
%--------
%---
%-
%
%
%-
%---
%--------
%-----------------
%-----------------------------------
%-----------------------------------------------------------------------

%=======================================================================
\section{Basic equations for matter surrounding a charged black hole}
\label{Basic_Eqs}
%=======================================================================

%=======================================================================
\subsection{General model}\label{gen_eq}
%=======================================================================
Following the Einstein cluster construction
\cite{Einstein:1939ms,Cardoso:2021wlq}, we begin by considering a single
species anisotropic charged fluid in spherical symmetry with vanishing
radial pressure, which one might imagine to be an effective fluid
description of a cloud of charged particles, each traveling on a
circular orbit, such that the averaged distribution is spherically
symmetric. The energy-momentum tensor is assumed to have the form
$T^{\mu}_{\nu}={\rm diag}(-\rho,0,\bar{P},\bar{P})$, where $\bar{P}$ is
the tangential pressure of the fluid. The effective energy-momentum
tensor may be rewritten in terms of the metric and appropriate vector
fields as:
%%%%%%%%%%%%%%%%%%%%%%%%%%%%%%%%%%%%%%%%%
\begin{equation}\label{Eq:EMTaniso}
    T^{\mu \nu} = \left(\rho + \bar{P}\right) u^{\mu} u^{\nu}
    + \bar{P} \left(
    g^{\mu \nu} - \hat{r}^{\mu} \hat{r}^{\nu}
    \right)~,
\end{equation}
%%%%%%%%%%%%%%%%%%%%%%%%%%%%%%%%%%%%%%%%%
where $u^{\mu}$ is the four-velocity of the fluid, which is assumed to
align with a timelike Killing vector field and satisfies the
normalization condition $u^{\mu}u_{\nu}=-1$. The quantity
$\hat{r}^{\mu}$ is a unit vector in the radial direction, which is
assumed to be orthogonal to the Killing vectors in the static and
spherically symmetric spacetime, within which we will work and is
normalized as $\hat{r}^{\mu}\hat{r}_{\mu}=1$.

Since the fluid is charged, we need to find out the associated field
equations governing the charge distribution as well. For this purpose,
we may use Maxwell's equations, having the following form (where we
define the field strength tensor as
$F_{\mu\nu}:=2\nabla_{[\mu}A_{\nu]}$):
%%%%%%%%%%%%%%%%%%%%%%%%%%%%%%%%%%%%%%%%%
\begin{equation}\label{Eq:MaxEq}
    \nabla_\nu F^{\mu \nu} = 4 \pi J^\mu ~,
\end{equation}
%%%%%%%%%%%%%%%%%%%%%%%%%%%%%%%%%%%%%%%%%
where $J^\mu := q n u^\mu$, with $q$ being the particle charge, and $n$
is an effective number density, defined so that $q n$ is the charge
density as viewed by observers coincident with $u^\mu$. Note that we do
not assume, $\rho=nm$, i.e., the density of charged particles need not
coincide with the density of the particles constituting the cloud, all
of which carry a mass $m$. Since $\rho$ and $n$ are assumed to be 
independent, one might regard $n$ as a quantity describing effective
charge overdensities for some underlying multispecies fluid description 
of the plasma.

The final ingredient is the Einstein's field equations, determining the
nature of the gravitational field,
%%%%%%%%%%%%%%%%%%%%%%%%%%%%%%%%%%%%%%%%%
\begin{equation}\label{Eq:GravEq}
G_{\mu \nu} = 8 \pi T_{\mu \nu}+
\frac{1}{2}\Big(4F{_\mu}{^\sigma}F_{\nu \sigma}
-g_{\mu \nu} F_{\sigma\tau}F^{\sigma\tau}\Big)\,.
\end{equation}
%%%%%%%%%%%%%%%%%%%%%%%%%%%%%%%%%%%%%%%%%
From the contracted Bianchi identity $\nabla_{\mu}G^{\mu}_{\nu}=0$, one
can show that the fluid must satisfy the following equation:
%%%%%%%%%%%%%%%%%%%%%%%%%%%%%%%%%%%%%%%%%
\begin{equation}\label{Eq:FlEq}
            \nabla_\nu T^{\mu \nu} = F^{\mu \nu} J_\nu~,
\end{equation}
%%%%%%%%%%%%%%%%%%%%%%%%%%%%%%%%%%%%%%%%%
where Eq.~\eqref{Eq:MaxEq} and the identity $\nabla_{[\sigma}F_{\mu
\nu]}=0$ have been employed. In addition, we also have the condition
that, $\nabla_{\mu}J^{\mu}=0$. The system of equations
(\ref{Eq:MaxEq}--\ref{Eq:FlEq}) can be closed by specifying an equation
of state relating $n$ to the other variables in the system, which we do
not yet impose for the sake of convenience and generality.

%=======================================================================
\subsection{Static and spherically symmetric spacetime} 
\label{Sec:Basic_Eqs_Sph}
%=======================================================================
We have spelled out the basic equations involving a charged cloud around
a BH, in a general form, in the above section. Here we
specialize to a static and spherically symmetric spacetime, whose line
element takes the following form:
%%%%%%%%%%%%%%%%%%%%%%%%%%%%%%%%%%%%%%%%%
\begin{equation}\label{Eq:LineElement}
ds^2 =  - f(r) dt^2 + \frac{dr^2}{1-2m(r)/r} 
            + r^2 d\Omega^2, 
\end{equation}
%%%%%%%%%%%%%%%%%%%%%%%%%%%%%%%%%%%%%%%%%
where $d\Omega^2 := d\theta^2+\sin^2\theta \, d\varphi^2$. Since the
above spacetime admits the existence of a timelike Killing vector
$(\partial/\partial t)^{\mu}$ and an angular Killing vector
$(\partial/\partial \varphi)^{\mu}$, one can identify the four-velocity
$u^\mu$ and the radial unit vector $\hat{r}^\mu$ with the following
ones:
%%%%%%%%%%%%%%%%%%%%%%%%%%%%%%%%%%%%%%%%%
\begin{equation}\label{Eq:Vectors}
\begin{aligned}
u^\mu =\frac{1}{\sqrt{f(r)}}\delta{_t}{^\mu}, \qquad 
\hat{r}^\mu = \delta{_r}{^\mu} \sqrt{1-\frac{2m(r)}{r}}~.
\end{aligned}
\end{equation}
%%%%%%%%%%%%%%%%%%%%%%%%%%%%%%%%%%%%%%%%%
For the vector potential, we assume:
%%%%%%%%%%%%%%%%%%%%%%%%%%%%%%%%%%%%%%%%%
\begin{equation}\label{Eq:VectorPotential}
        A_\mu = {\delta{_\mu}{^t}} \phi(r)~.
\end{equation}
%%%%%%%%%%%%%%%%%%%%%%%%%%%%%%%%%%%%%%%%%
where the scalar potential $\phi$ depends on the radial coordinate
alone, due to spherical symmetry of the background spacetime. The
spherical symmetry of the system also results in the following equation
$\nabla_{\mu}J^{\mu}=0$ to be identically satisfied. The Einstein's
equations, in particular the $(0,0)$, $(r,r)$ and $(\theta,\theta)$
equations, yield
%%%%%%%%%%%%%%%%%%%%%%%%%%%%%%%%%%%%%%%%%
\begin{equation}\label{Eq:XEFE1}
    Y'(r) = \frac{8 \pi  r^2 \rho (r)}{r-2 m(r)}~,
\end{equation}
%%%%%%%%%%%%%%%%%%%%%%%%%%%%%%%%%%%%%%%%%
%%%%%%%%%%%%%%%%%%%%%%%%%%%%%%%%%%%%%%%%%
\begin{equation}\label{Eq:XEFE2}
    2m'(r)=8 \pi  r^2 \left[\rho (r)+\psi(r)\right]~,
\end{equation}
%%%%%%%%%%%%%%%%%%%%%%%%%%%%%%%%%%%%%%%%%
and, finally,
%%%%%%%%%%%%%%%%%%%%%%%%%%%%%%%%%%%%%%%%%
\begin{equation}\label{Eq:XFl3}
\bar{P}(r)= \frac{\left[ \rho(r) f'(r) - 2 q \sqrt{f(r)} n(r) \phi'(r) \right]r}{4 f(r)}~.
\end{equation}
%%%%%%%%%%%%%%%%%%%%%%%%%%%%%%%%%%%%%%%%%
The $(\varphi,\varphi)$ component of the Einstein's equations will yield
an identical expression for the tangential pressure $\bar{P}(r)$ and
hence adds nothing new to the above discussion. The quantities $\psi(r)$
and $Y(r)$ are defined for convenience, having the following form:
%%%%%%%%%%%%%%%%%%%%%%%%%%%%%%%%%%%%%%%%%
\begin{equation}\label{Eq:PsiDef}
\psi(r):=\frac{\phi '(r)^2}{8 \pi f(r)} \left(1-\frac{2 m(r)}{r}\right)~,
\end{equation}
%%%%%%%%%%%%%%%%%%%%%%%%%%%%%%%%%%%%%%%%%
and the quantity $Y(r)$ is defined by the expression:
%%%%%%%%%%%%%%%%%%%%%%%%%%%%%%%%%%%%%%%%%
\begin{equation}\label{Eq:YDef}
Y(r)-Y_{\infty}=\ln \left(\frac{r f(r)}{r-2 m(r)}\right)~,
\end{equation}
%%%%%%%%%%%%%%%%%%%%%%%%%%%%%%%%%%%%%%%%%
where $Y_{\infty}$ corresponds to the value of the function $Y(r)$ at
infinity and it ensures that, at large distances the metric function
$f(r)$ has the same functional form (up to a constant factor) as the
metric function $g^{rr}$ and the appropriate flat limit for both of
these metric functions can be obtained.

Moreover, the surface $r=2m(r)$ represents a null surface, since its
normal vector corresponds to $\ell_{\mu}=(1-2m')\delta_{\mu}^{r}$, such that,
$\ell_{\mu}\ell^{\mu}=(1-2m')^{2}g^{rr}$, which identically vanishes on
the $r=2m(r)$ surface. Also, the energy density $\rho$ must vanish on
this surface so that $Y'(r)$ remains finite on the same. This in turn
implies from Eq. \eqref{Eq:YDef} that $f(r)$ and $r-2m(r)$ must have
coinciding zeros. Therefore, on the surface $r=2m(r)$, the metric
function $f(r)=(\partial/\partial t)^{\mu}(\partial/\partial t)_{\mu}$
must vanish as well. Therefore, the largest root of this equation
$r=2m(r)$, denoted by $r_{+}$, corresponds to an event horizon for this
spacetime. Further, for $r>r_{+}$, the equation for $Y(r)$ presented in
Eq. \eqref{Eq:XEFE1} shows that $Y'(r)>0$, owing to the weak energy
condition, ensuring $\rho(r)>0$. As a consequence the function $Y(r)$
has a monotonic behavior beyond the surface $r=r_{+}$. This fact will
be of importance later.  

Also note that $\psi(r)$ represents the energy density in the
electromagnetic field of the charged cloud and $Y(r)$ quantifies the
difference between the $g_{tt}$ and the $g^{rr}$ components of the
metric tensor. Further, Eq.~\eqref{Eq:XFl3} is also equivalent to the
conservation relation in Eq.~\eqref{Eq:FlEq}; thus, we need not consider
the conservation equation once again. The only remaining equation
corresponds to the Maxwell's equation in the context of static and
spherically symmetric background spacetime, which reads
%%%%%%%%%%%%%%%%%%%%%%%%%%%%%%%%%%%%%%%%%
\begin{equation}\label{Eq:ddphi}
\begin{aligned}
\phi ''(r)=&\frac{1}{2 r (r-2 m(r))} \biggl[8 \pi q r^2 n(r) \sqrt{f(r)} \\
& \qquad 
+\phi '(r) \left(8 \pi r^3 \rho (r) + 8 m(r) - 4 r \right)
\biggr] ~.
\end{aligned}
\end{equation}
%%%%%%%%%%%%%%%%%%%%%%%%%%%%%%%%%%%%%%%%%
We now have all the equations that gravity and electromagnetism have to
offer. However, the system of equations is not closed, since the
Eqs.~\eqref{Eq:XEFE1} ---\eqref{Eq:ddphi} constitute a system of four
independent equations for the six variables $Y(r)$, $m(r)$,
$\bar{P}(r)$, $\rho(r)$, $n(r)$, and $\phi(r)$. Thus we need additional
supplementary conditions, which we discuss in the next sections. 

Before proceeding further, it is perhaps appropriate to briefly consider
in this framework the emergence of the Reissner-Nordstr\"om geometry, and
for that purpose it will be interesting to study electrovacuum solution
from the above equations. Consider a region of spacetime ($r_{+}<r<R$)
with no matter field, but with an electromagnetic field being present
due to the BH being charged. Since there is no matter distribution, it
is apt to set $\rho=0$ in this region and hence from Eq.
\eqref{Eq:XEFE1}, it follows that $Y(r)=Y_{0}$ is a constant in that
region. However, let there be some spherically symmetric matter
distribution away from this region ($r>R$); $Y'(r)$ will be nonzero
there, and from the monotonic behavior for $Y(r)$, it follows that
$Y_{\infty}>Y_{0}$. Thus it follows from Eq. \eqref{Eq:YDef} that the
$(-g_{tt})$ component of the metric function reads
%%%%%%%%%%%%%%%%%%%%%%%%%%%%%%%%%%%%%%%%%
\begin{align}\label{Eq:evacsol}
f_{\rm EV}(r)&=e^{\Delta Y_{0}}\left(1-\frac{2 m(r)}{r}\right)
\nonumber
\\
&=e^{\Delta Y_{0}}\left( 1-\frac{2M_{\rm BH}}{r}\right)+ \frac{Q_{\rm BH}^2}{r^2}\,,
\end{align}
%%%%%%%%%%%%%%%%%%%%%%%%%%%%%%%%%%%%%%%%%
where we have defined $\Delta Y_{0}:= Y_{0}-Y_{\infty}$, which is a
negative quantity. In order to determine the mass function, we use the
result that the electrostatic potential reads in the vacuum region
outside the event horizon, as $\phi(r)=(Q_{\rm BH}/r)$, and hence the
electrostatic energy yields $\psi(r)=(Q_{\rm BH}^{2}/8 \pi
r^{4})e^{-\Delta Y_{0}}$, from Eq. \eqref{Eq:PsiDef}. Therefore,
subsequent integration of Eq. \eqref{Eq:XEFE2} provides the mass
function $m(r)$, which reads $m(r)=M_{\rm BH}-( Q_{\rm
BH}^{2}/2r)e^{-\Delta Y_{0}}$, whose substitution provides the final
expression in Eq. \eqref{Eq:evacsol}. Note that the extremal limit
corresponds to $M_{\rm BH}=e^{-\Delta Y_{0}}Q_{\rm BH}$. However, if
the spacetime has no matter density outside the event horizon, then
$Y_{0}$ and $Y_{\infty}$ will coincide and hence $\Delta Y_{0}$ will
vanish. In which case, the Reissner-Nordstr\"{o}m expression will be
obtained. To recover the standard Reissner-Nordstr\"{o}m or
Schwarzschild expression at large $r$, we henceforth choose
$Y_{\infty}=0$, so that $\Delta Y_0 = Y_0$. In what follows, we will
adopt this strategy to parametrize a charged BH surrounded by a shell of
charged matter with compact support in the radial direction. 

%-----------------------------------------------------------------------
%-----------------------------------
%-----------------
%--------
%---
%-
%
%
%-
%---
%--------
%-----------------
%-----------------------------------
%-----------------------------------------------------------------------

%=======================================================================
\section{Debye model}\label{Sec:Debye}
%=======================================================================

%%%%%%%%%%%%%%%%%%%%%%%%
%%%%%%%%%%%%%%%%%%%%%%%%
\begin{figure*}
    \centering
    \includegraphics[scale=0.5]{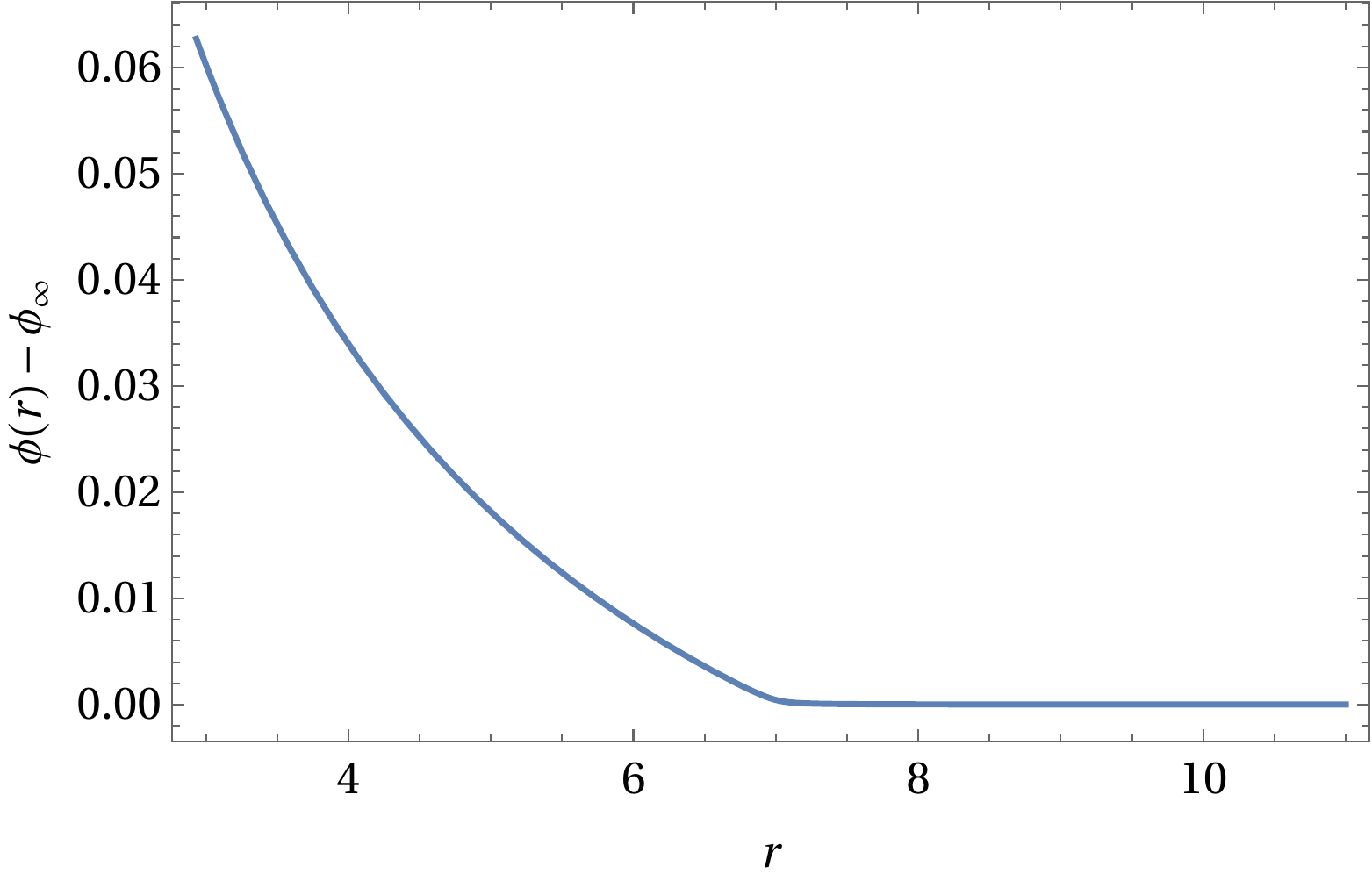}
    \qquad 
    \includegraphics[scale=0.465]{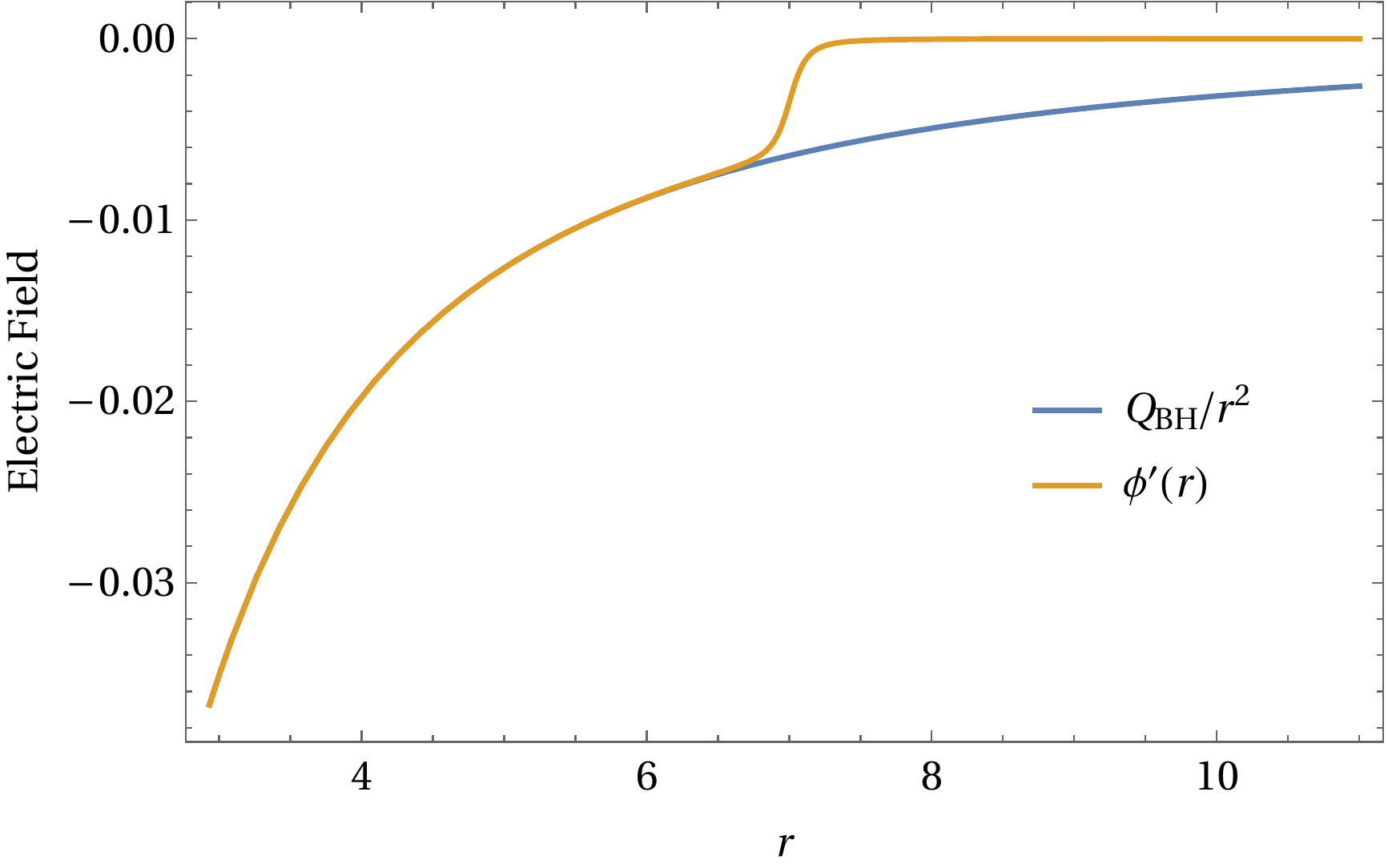}
    \caption{Numerical solutions of Eq. \eqref{Eq:ddphi2} for $H(x)=x$,
    $Y(r)$ given by Eq. \eqref{Eq:Ystep} and \eqref{Eq:stepfunction},
    with boundary data $\phi(r_{\rm max})=1$, $\phi'(r_{\rm max})=0$.
    Here, we use exaggerated parameter choices: $\nu_0=1/10$, $Y_0 =
    -0.5122$, $r_0=7$, $\lambda_0=1/8$, and $m_R=2.450$. The parameters
    have been adjusted so that the resulting BH mass parameter is
    $M_{\rm BH}=1$. The charge parameter is $Q_{\rm BH}=0.3159$ and the
    ADM mass is $M_{\rm ADM}=2$. The plot on the left is the electric
    potential $\phi(r)$, and the plot on the right is the electric field
    $\phi'(r)$, and the unshielded electric field profile $-Q_{\rm
    BH}/r^2$.}
    \label{fig:ShieldingSolutionPhiE}
\end{figure*}
%%%%%%%%%%%%%%%%%%%%%%%%
%%%%%%%%%%%%%%%%%%%%%%%%

%%%%%%%%%%%%%%%%%%%%%%%%
%%%%%%%%%%%%%%%%%%%%%%%%
\begin{figure*}
    \centering
    \includegraphics[scale=0.45]{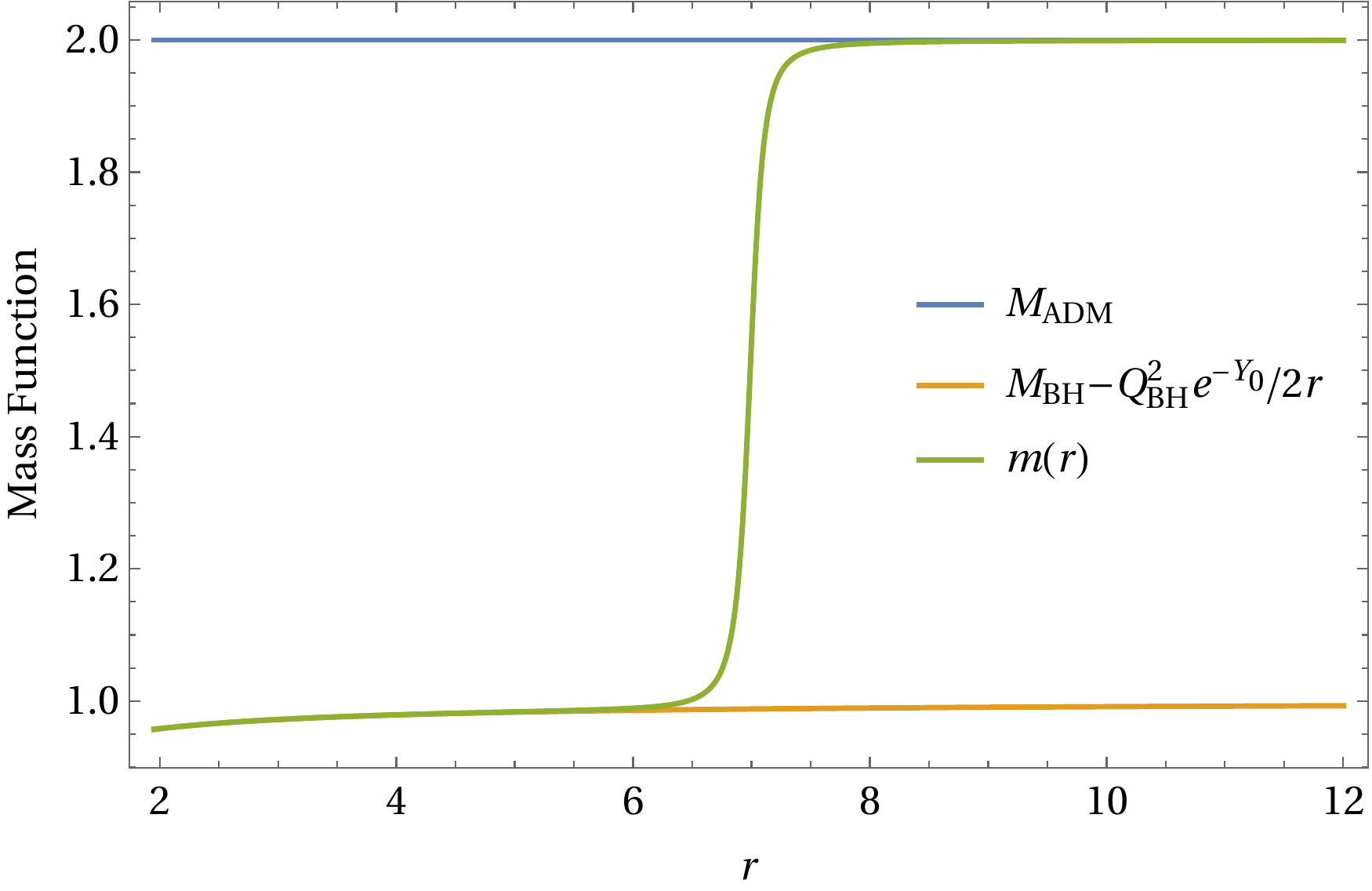}
    \qquad
    \includegraphics[scale=0.45]{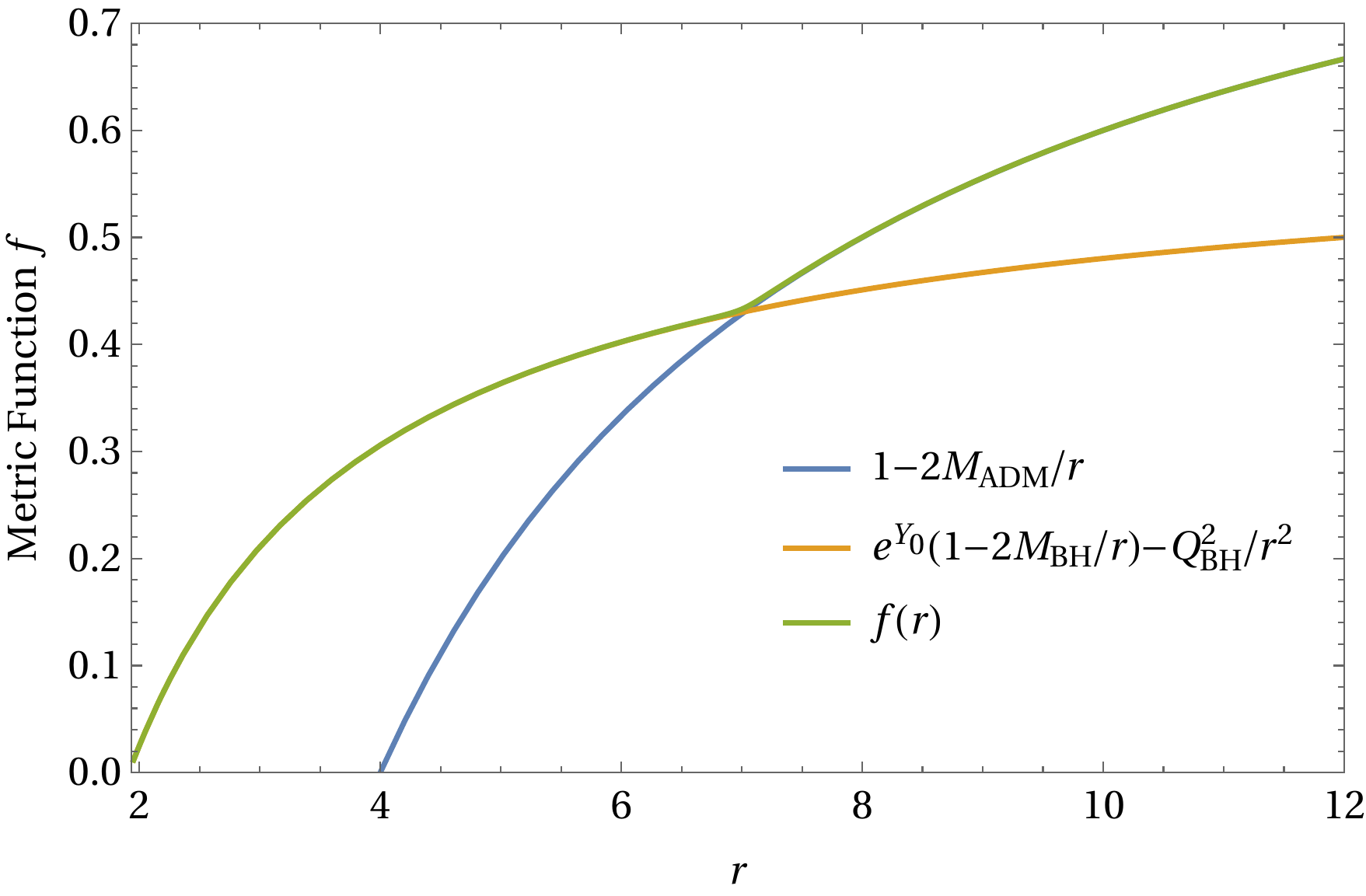}
    \caption{Numerical solutions of Eqs. \eqref{Eq:mpEq} and
    \eqref{Eq:fpEq} corresponding to the electric potential solution of
    Fig. \ref{fig:ShieldingSolutionPhiE}, with BH charge parameter
    $Q_{\rm BH}=0.3159$, mass parameter $M_{\rm BH}=1$, and the ADM mass
    $M_{\rm ADM}=6$. On the left is $m(r)$, plotted against the value of
    $m(\infty)$ and unshielded mass function $(M_{\rm BH}-{Q_{\rm BH}^2
    e^{-Y_0}}/{2 r})$. On the right is the metric component $f(r)$,
    plotted against the Schwarzschild value $1-2m(\infty)/r$ and the
    unshielded Reissner-Nordstr\"{o}m metric component given by Eq.
    \eqref{Eq:evacsol}.} 
    \label{fig:ShieldingSolutionmfE}
\end{figure*}
%%%%%%%%%%%%%%%%%%%%%%%%
%%%%%%%%%%%%%%%%%%%%%%%%

%%%%%%%%%%%%%%%%%%%%%%%%
%%%%%%%%%%%%%%%%%%%%%%%%

\begin{figure*}
    \centering
    \includegraphics[scale=0.45]{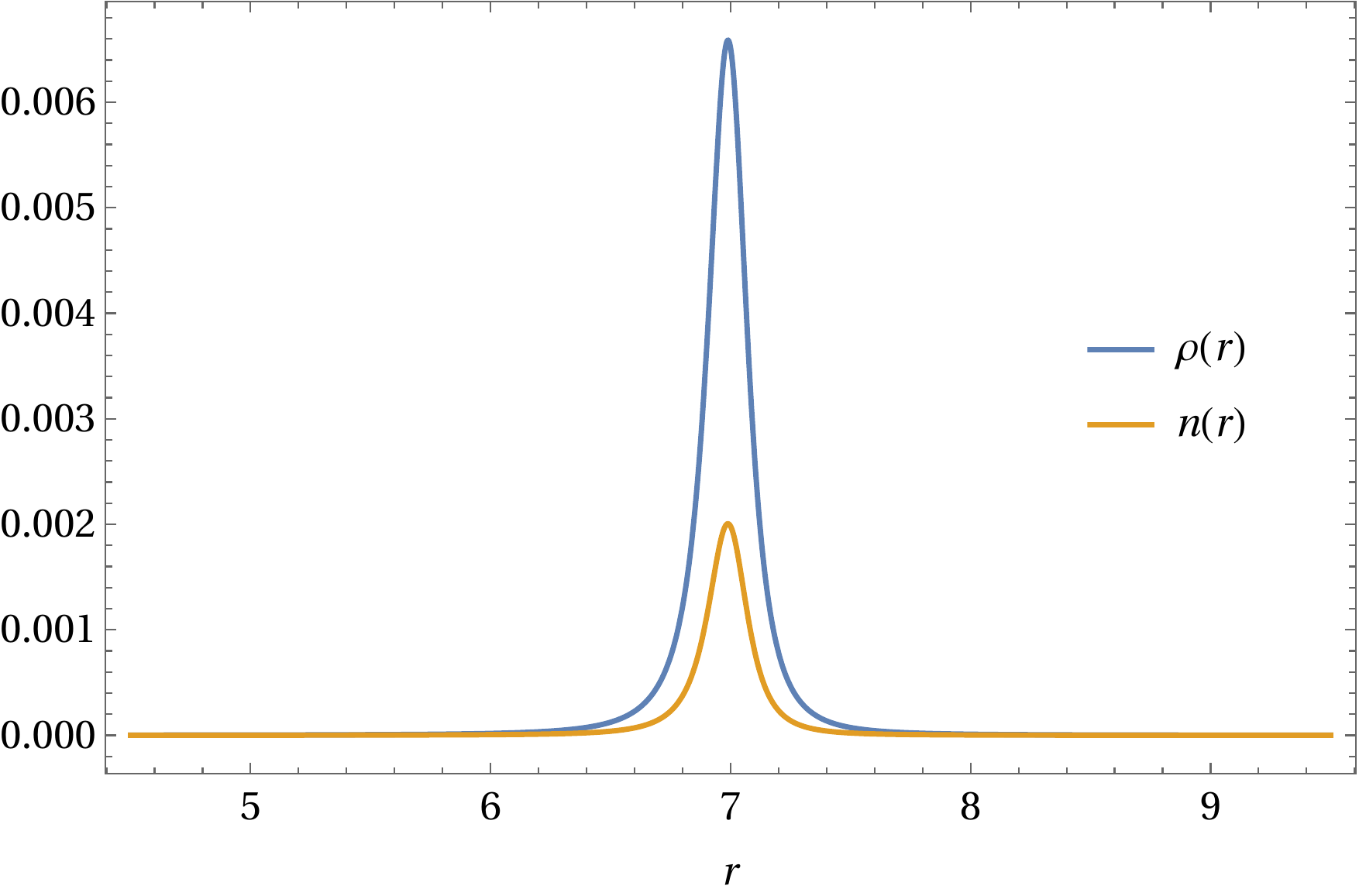}
    \qquad
    \includegraphics[scale=0.49]{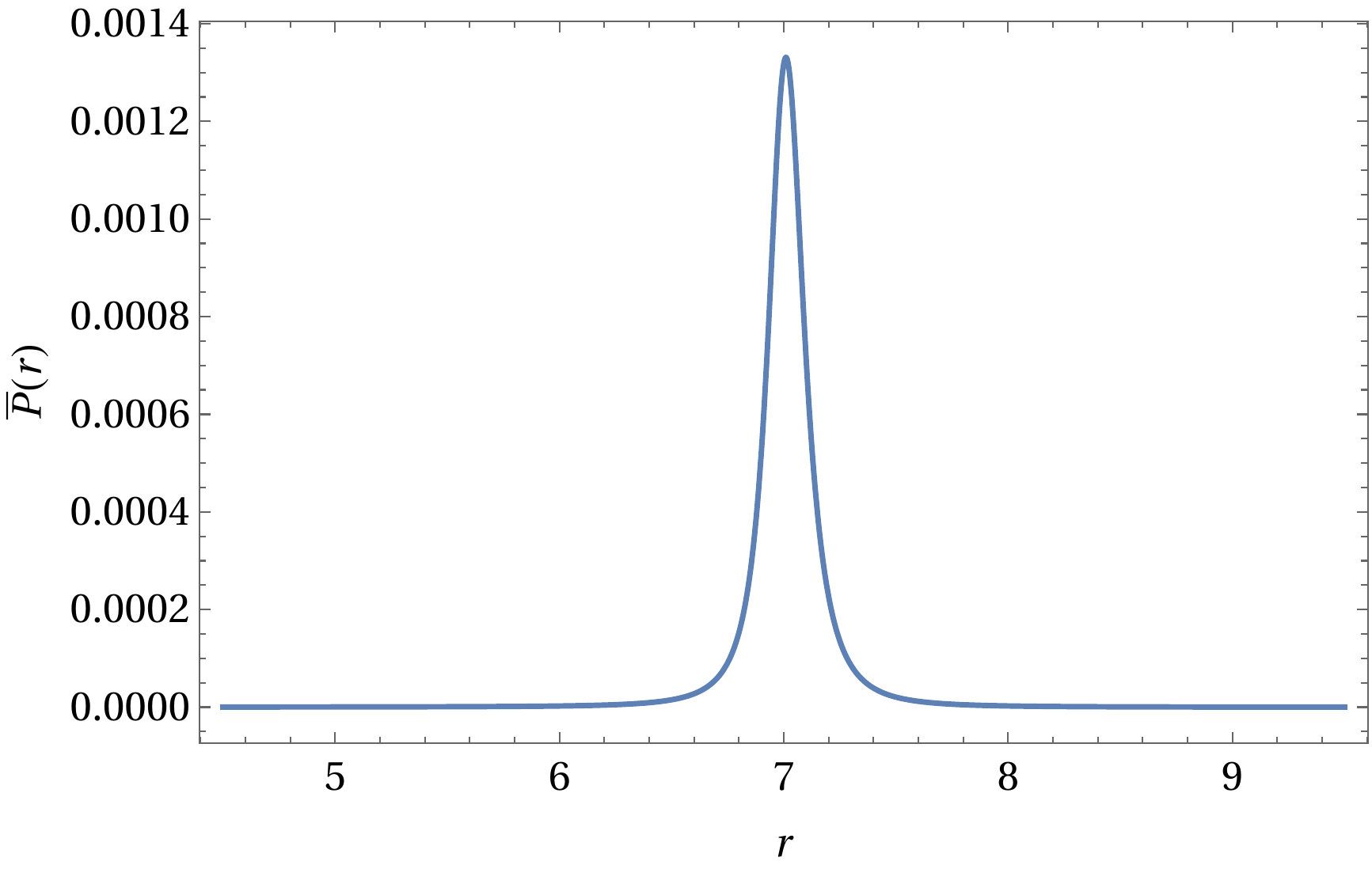}
    \caption{Density and tangential stress profiles corresponding to the
    solutions in Figs. \ref{fig:ShieldingSolutionPhiE} and
    \ref{fig:ShieldingSolutionmfE}. On the left are the energy density
    profile $\rho(r)$ and the charge density profile $n(r)$, which are
    nearly identical up to a scaling factor. On the right is the
    tangential stress profile $\bar{P}$; the solution satisfies the
    dominant energy condition. Qualitatively similar plots can be
    obtained for other parameter choices.}
    \label{fig:DensityE}
\end{figure*}
%%%%%%%%%%%%%%%%%%%%%%%%
%%%%%%%%%%%%%%%%%%%%%%%%

In this section, we will solve Eq. \eqref{Eq:ddphi} and show that the
screening mechanism arises naturally in a limit for a rather large class
of physically motivated charge density profiles. As mentioned in the
preceding section, one has a system of four equations for six
variables---two supplementary equations are required. Here, we supply
these conditions by choosing $Y'(r)$ and a relation between the charge
density $qn(r)$ and the electrostatic potential $\phi(r)$, motivated by
the Debye shielding model in plasma physics. 

In the Debye shielding model \cite{Chen2012,Bittencourt2013}, one
assumes that the number densities for each particle species satisfies a
Maxwell-Boltzmann distribution in the particle energies, and upon
integrating out the particle momenta, one finds that the distribution is
a function of the electrostatic potential. In the high-temperature
limit, one recovers a charge density proportional to the electrostatic
potential. In curved spacetime, one should note that since the charge
density is defined with respect to the fluid four-velocity, the
electrostatic potential in the comoving frame has the form $A_\mu
u^\mu=\phi/\sqrt{f}$ (the rhs being the expression in spherical
symmetry). If we also require that the charge density is a fraction of
the matter density $\rho(r)$, one arrives at the expression:
\begin{equation}\label{Eq:numberdensityapprox}
qn(r) \approx \nu_0 \frac{\phi(r)\rho(r)}{\sqrt{f(r)}} ,
\end{equation}
%%%%%%%%%%%%%%%%%%%%%%%%%%%%%%%%%%%%%%%%%
which one may regard as the leading order term in the high $k T$
expansion of the general expression:
%%%%%%%%%%%%%%%%%%%%%%%%%%%%%%%%%%%%%%%%%
\begin{equation}\label{Eq:numberdensitygen}
qn(r)=
    \frac{\nu_0 k T}{\bar{q}}\frac{\rho(r)}
    {\sqrt{f(r)}} H(\bar{q}\phi(r)/kT)~,
\end{equation}
%%%%%%%%%%%%%%%%%%%%%%%%%%%%%%%%%%%%%%%%%
where $H(\cdot)$ is a suitably chosen distribution function of the
energies of the individual charged particles and $\bar{q}$ is some
arbitrary test charge. One might, following the Debye shielding model,
construct $H(\cdot)$ from the Maxwell-Boltzmann distribution, so that
$H(x)=\exp(x)-1$, but Eq. \eqref{Eq:numberdensitygen} should apply for
any distribution function that reduces to a homogeneous linear function
in the high temperature limit for an appropriate rescaling of $\nu_0$. 
%A Maxwell-Boltzmann-like distribution may be preferred, since a
%constant shift in the potential can in the high $k T$ limit be absorbed
%into $\nu_0$. 
With Eqs. \eqref{Eq:numberdensitygen} and \eqref{Eq:XEFE1}, Eq.
\eqref{Eq:ddphi} has the general form:
%%%%%%%%%%%%%%%%%%%%%%%%%%%%%%%%%%%%%%%%%
\begin{equation}\label{Eq:ddphi2}
\phi ''(r)+\frac{2 \phi '(r)}{r}
=
    \frac{1}{2} Y'(r) 
    \left[
        \frac{\nu_0 kT q H(\bar{q}\phi(r)/kT)}{\bar{q} r}+\phi '(r)
    \right]
\end{equation}
%%%%%%%%%%%%%%%%%%%%%%%%%%%%%%%%%%%%%%%%%
\noindent so that, for a choice of distribution function $H(\cdot)$, one
can solve for $\phi(r)$, given a function $Y(r)$.

\begin{figure*}
    \centering
    \includegraphics[scale=0.5]{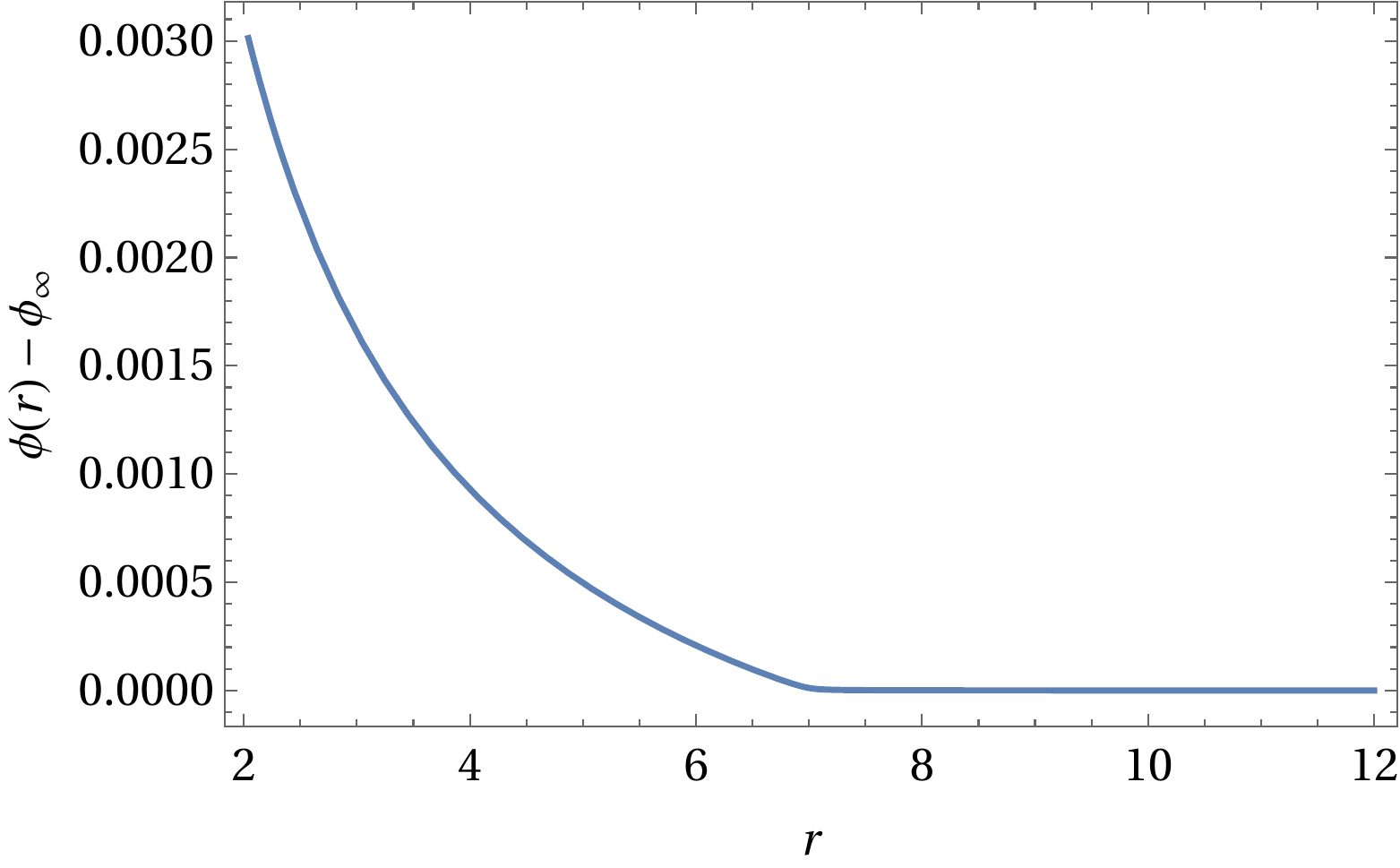}
    \qquad 
    \includegraphics[scale=0.465]{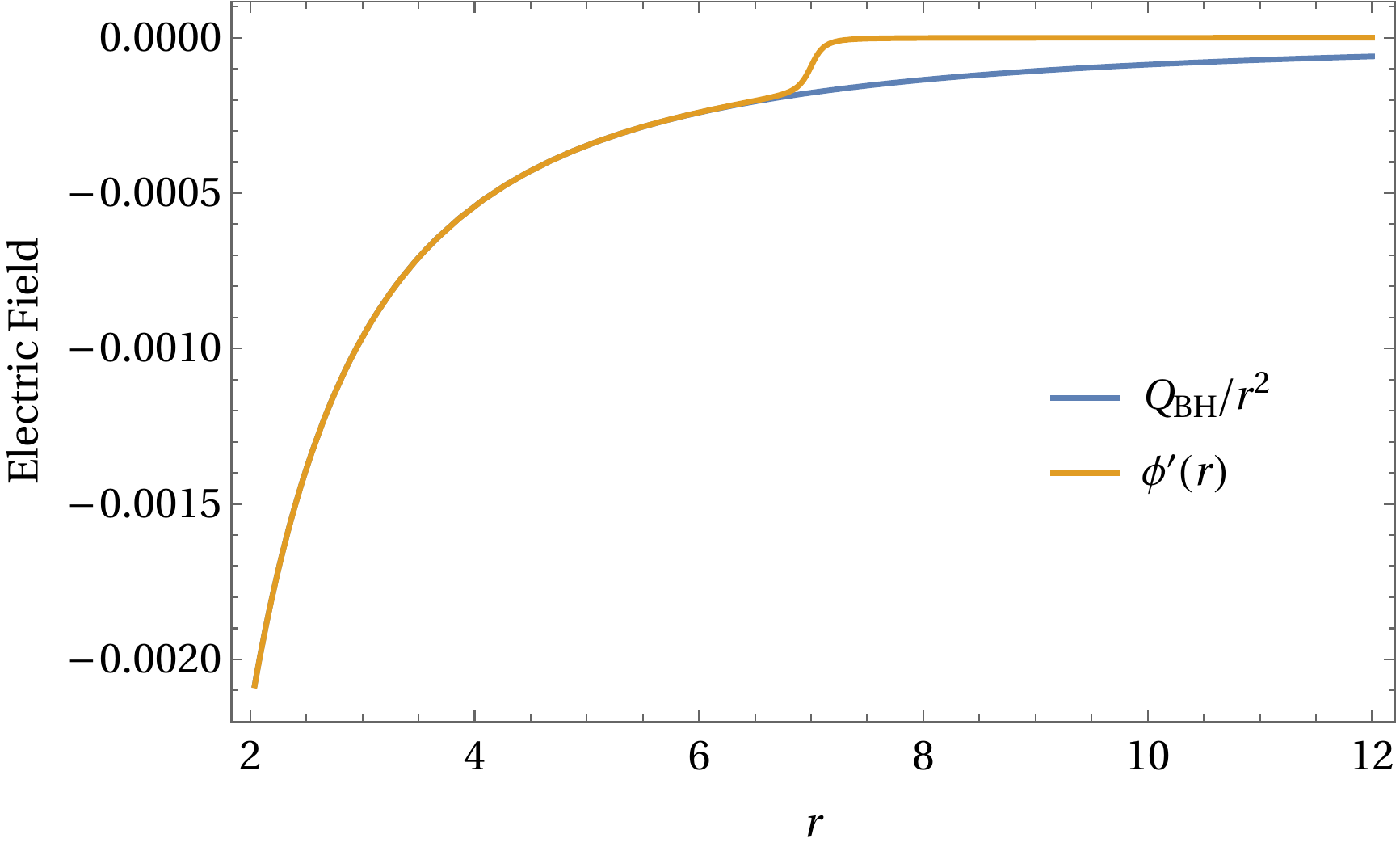}
    \caption{Numerical solutions of Eq. \eqref{Eq:ddphi2} for $H(x)=x$,
    $Y(r)$ given by Eq. \eqref{Eq:Ystep} and \eqref{Eq:stepfunction},
    with boundary data $\phi(r_{\rm max})=1$, $\phi'(r_{\rm max})=0$.
    Here, we use realistic parameter choices: $\nu_0=1/50$, $Y_0 =
    -0.1276$, $r_0=7$, $\lambda_0=0.125$, and $m_R=108.2$. The
    parameters have been adjusted so that the resulting BH mass
    parameter is $M_{\rm BH}=1$. The BH charge parameter is $Q_{\rm
    BH}=8.667\times10^{-3}$ and the ADM mass is $M_{\rm ADM}=1.3$. The
    plot on the left is the electric potential $\phi(r)$, and the plot
    on the right is the electric field $\phi'(r)$ (in yellow), and the
    unshielded electric field profile $-Q_{\rm BH}/r^2$ (in blue).}
    \label{fig:ShieldingSolutionPhiR}
\end{figure*}

\begin{figure*}
    \centering
    \includegraphics[scale=0.45]{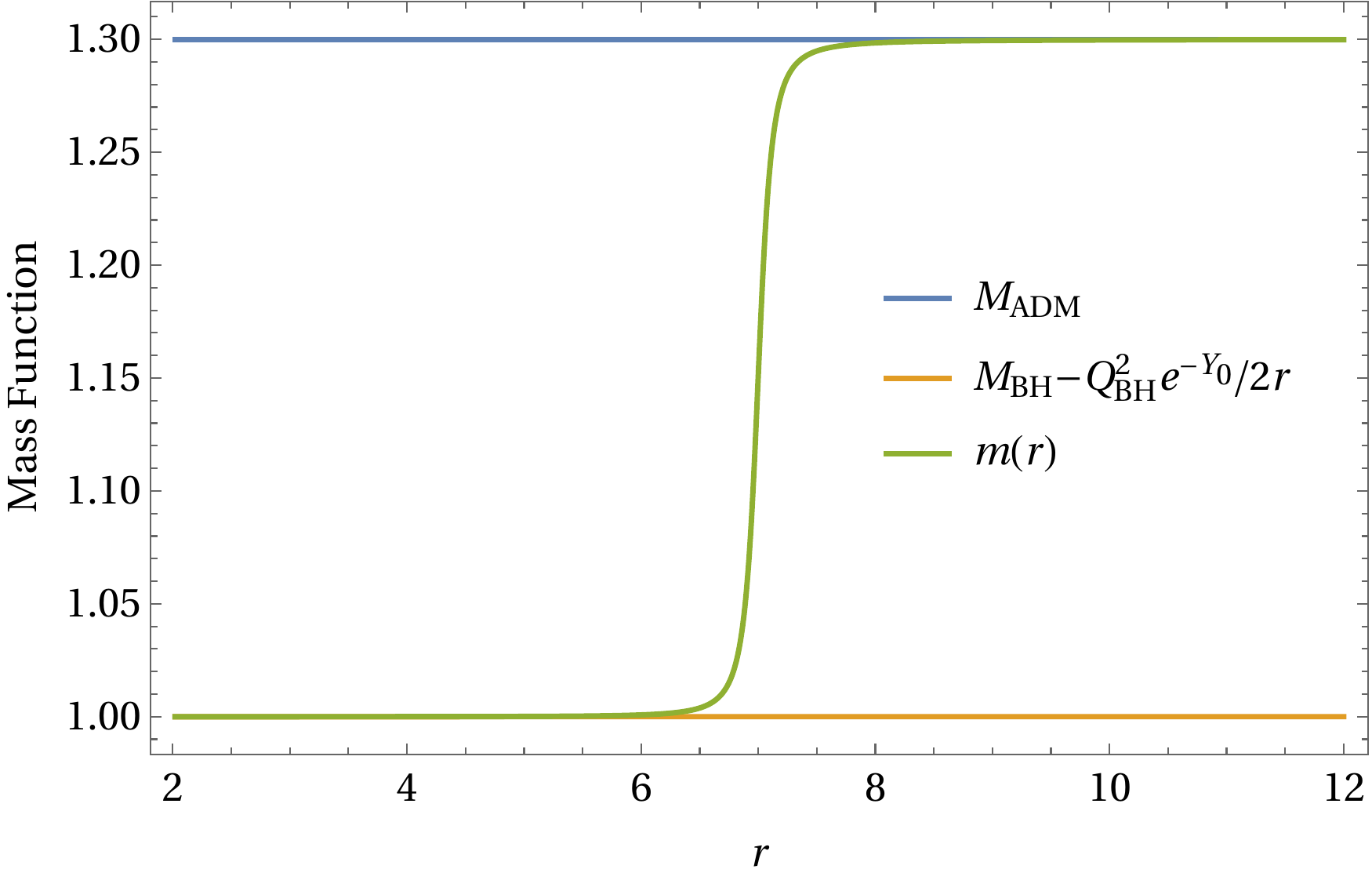}
    \qquad
    \includegraphics[scale=0.45]{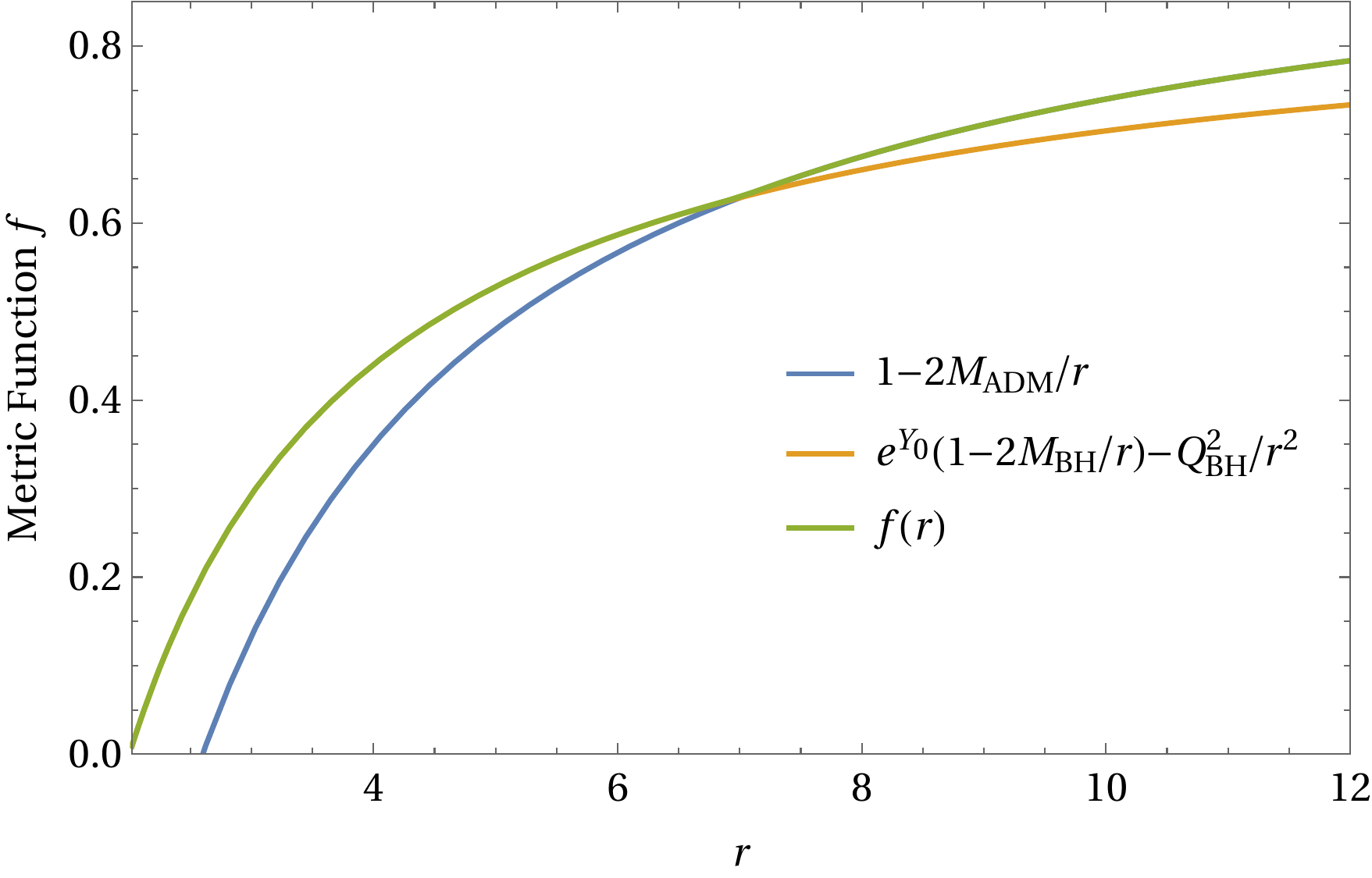}
    \caption{Numerical solutions of Eqs. \eqref{Eq:mpEq} and
    \eqref{Eq:fpEq} corresponding to the electric potential solution of
    Fig. \ref{fig:ShieldingSolutionPhiR}, with BH charge parameter
    $Q_{\rm BH}=8.667\times10^{-3}$, mass parameter $M_{\rm BH}=1$, and
    the ADM mass $M_{\rm ADM}=1.3$. On the left is $m(r)$ (in green),
    plotted against the value of $m(\infty)$ (in blue) and unshielded
    mass function $(M_{\rm BH}-{Q_{\rm BH}^2 e^{-Y_0}}/{2 r})$ (in
    yellow). On the right is the metric component $f(r)$ (in green),
    plotted against the Schwarzschild value $1-2m(\infty)/r$ (in blue)
    and the unshielded Reissner-Nordstr\"{o}m metric component given by Eq.
    \eqref{Eq:evacsol} (in yellow).} 
    \label{fig:ShieldingSolutionmfR}
\end{figure*}

One can motivate choices for $Y(r)$ from Eqs. \eqref{Eq:XEFE1} and
\eqref{Eq:YDef}, and the condition that outside the event horizon,
$Y(r)$ is a monotonic function. As a consequence, the most general
spherically symmetric solution to the Einstein-Maxwell system in the
absence of local charge and energy density is given by Eq.
\eqref{Eq:evacsol}. Now the system we consider is a charged BH
surrounded by charged matter fields with compact support; if we assume a
(electro)vacuum near the horizon and at large $r$, then the function
$Y(r)$ is a constant near the horizon and at large $r$, but these
constant values $Y_{0}$ and $Y_{\infty}=0$ will be different, with
$Y_{0}<Y_{\infty}=0$. Therefore, it follows that $\Delta Y_0=Y_0<0$ near
the horizon. These considerations suggest that $Y(r)$ has a step
function profile:
%%%%%%%%%%%%%%%%%%%%%%%%%%%%%%%%%%%%%%%%%
\begin{equation}\label{Eq:Ystep}
Y(r) = Y_0 \,
\sigma\left(\frac{r-r_0}{\lambda_0}\right)~,
\end{equation}
%%%%%%%%%%%%%%%%%%%%%%%%%%%%%%%%%%%%%%%%%
where, $\sigma(x)=0$ for $x \gg 1$ and $\sigma(x)=-1$ for $x \ll -1$. A
particularly simple choice, which we employ in our numerical solutions,
is the following:
%%%%%%%%%%%%%%%%%%%%%%%%%%%%%%%%%%%%%%%%%
\begin{equation}\label{Eq:stepfunction}
\sigma(x)=\frac{1}{2}\left[\frac{x}{\sqrt{x^2+1}}-1\right]~.
\end{equation}
%%%%%%%%%%%%%%%%%%%%%%%%%%%%%%%%%%%%%%%%%
We obtain numerical solutions for \eqref{Eq:ddphi2} with $H(x)=x$ [to
recover Eq. \eqref{Eq:numberdensityapprox} in appropriate limit)] with
the choice of $Y$ in Eqs. \eqref{Eq:Ystep} and \eqref{Eq:stepfunction},
and for some choice of the constant $Y_0$. We integrate Eq.
\eqref{Eq:ddphi2} inward from an initial point $r_{\rm max} \gg r_{+}$,
with initial data $\phi(r_{\rm max})=\phi_0 \neq 0$, $\phi'(r_{\rm
max})=0$. As illustrated in Figs. \ref{fig:ShieldingSolutionPhiE} and
\ref{fig:ShieldingSolutionPhiR}, the result is consistent with what one
might expect; for $r \ll r_0$, the potential has the form $\phi(r) =
\phi_0 + Q_{\rm BH}/{r}$, and for $r \gg r_0$ $\phi(r) = \phi_0$. The BH
charge $Q_{\rm BH}$ is obtained by fitting the portion of the solution
near the horizon to $\phi'(r) = -Q_{\rm BH}/{r^2}$. The BH mass, on the
other hand, is not fixed by the solution, but one can specify the mass
to charge ratio by way of the dimensionless parameter $m_R$:
%%%%%%%%%%%%%%%%%%%%%%%%%%%%%%%%%%%%%%%%%
\begin{equation}\label{Eq:mass2charge}
\frac{M_{\rm BH}}{Q_{\rm BH}} = m_R e^{-Y_0/2} ~,
\end{equation}
%%%%%%%%%%%%%%%%%%%%%%%%%%%%%%%%%%%%%%%%%
where, we require $m_R \geq 1$, with equality corresponding to the
extremal limit, also see Eq. \eqref{Eq:evacsol}. Given this form for
$Y(r)$ and the potential $\phi(r)$ solved from Eq. \eqref{Eq:ddphi2},
both the mass function $m(r)$ and the metric component $f(r)$ can be
derived. First of all, one solves the following equation using the
solution for $\phi(r)$:
%%%%%%%%%%%%%%%%%%%%%%%%%%%%%%%%%%%%%%%%%
\begin{equation}\label{Eq:mpEq}
2m'(r)=[r-2 m(r)]Y'(r)+r^2 e^{-Y(r)} \phi '(r)^2~,
\end{equation}
%%%%%%%%%%%%%%%%%%%%%%%%%%%%%%%%%%%%%%%%%
and then using the solution for the mass function from the above
equation, to solve 
%%%%%%%%%%%%%%%%%%%%%%%%%%%%%%%%%%%%%%%%%
\begin{equation}\label{Eq:fpEq}
\frac{f'(r)}{f(r)}=\frac{2 m(r)-2 r m'(r)}{r^2-2 r m(r)}+Y'(r)~.
\end{equation}
%%%%%%%%%%%%%%%%%%%%%%%%%%%%%%%%%%%%%%%%%
We would like to emphasize that both Eqs. \eqref{Eq:mpEq} and
\eqref{Eq:fpEq} are a straightforward rewriting of Eqs. \eqref{Eq:XEFE2}
and \eqref{Eq:XEFE1}, respectively. The solutions of the above
differential equations are illustrated in Figs.
\ref{fig:ShieldingSolutionmfE} and \ref{fig:ShieldingSolutionmfR}. The
corresponding energy density, charge density, and tangential pressure for
an exaggerated parameter choice is displayed in Fig.
\ref{fig:DensityE}. We choose the parameters so that the energy density 
vanishes roughly below the Innermost Stable Circular Orbit (ISCO) for 
uncharged particles surrounding a slightly charged BH.

%%%%%%%%%%%%%%%%%%%%%%%%
%%%%%%%%%%%%%%%%%%%%%%%%

%%%%%%%%%%%%%%%%%%%%%%%%
%%%%%%%%%%%%%%%%%%%%%%%%
%%%%%%%%%%%%%%%%%%%%%%%%
%%%%%%%%%%%%%%%%%%%%%%%%

%%%%%%%%%%%%%%%%%%%%%%%%
%%%%%%%%%%%%%%%%%%%%%%%%
%%%%%%%%%%%%%%%%%%%%%%%%
%%%%%%%%%%%%%%%%%%%%%%%%

%-----------------------------------------------------------------------
%-----------------------------------
%-----------------
%--------
%---
%-
%
%
%-
%---
%--------
%-----------------
%-----------------------------------
%-----------------------------------------------------------------------

%=======================================================================
\section{Black hole parametrization}\label{Sec:BH_param}
%=======================================================================

%%%%%%%%%%%%%%%%%%%%%%%%
%%%%%%%%%%%%%%%%%%%%%%%%
\begin{figure*}
    \centering
    \includegraphics[scale=0.45]{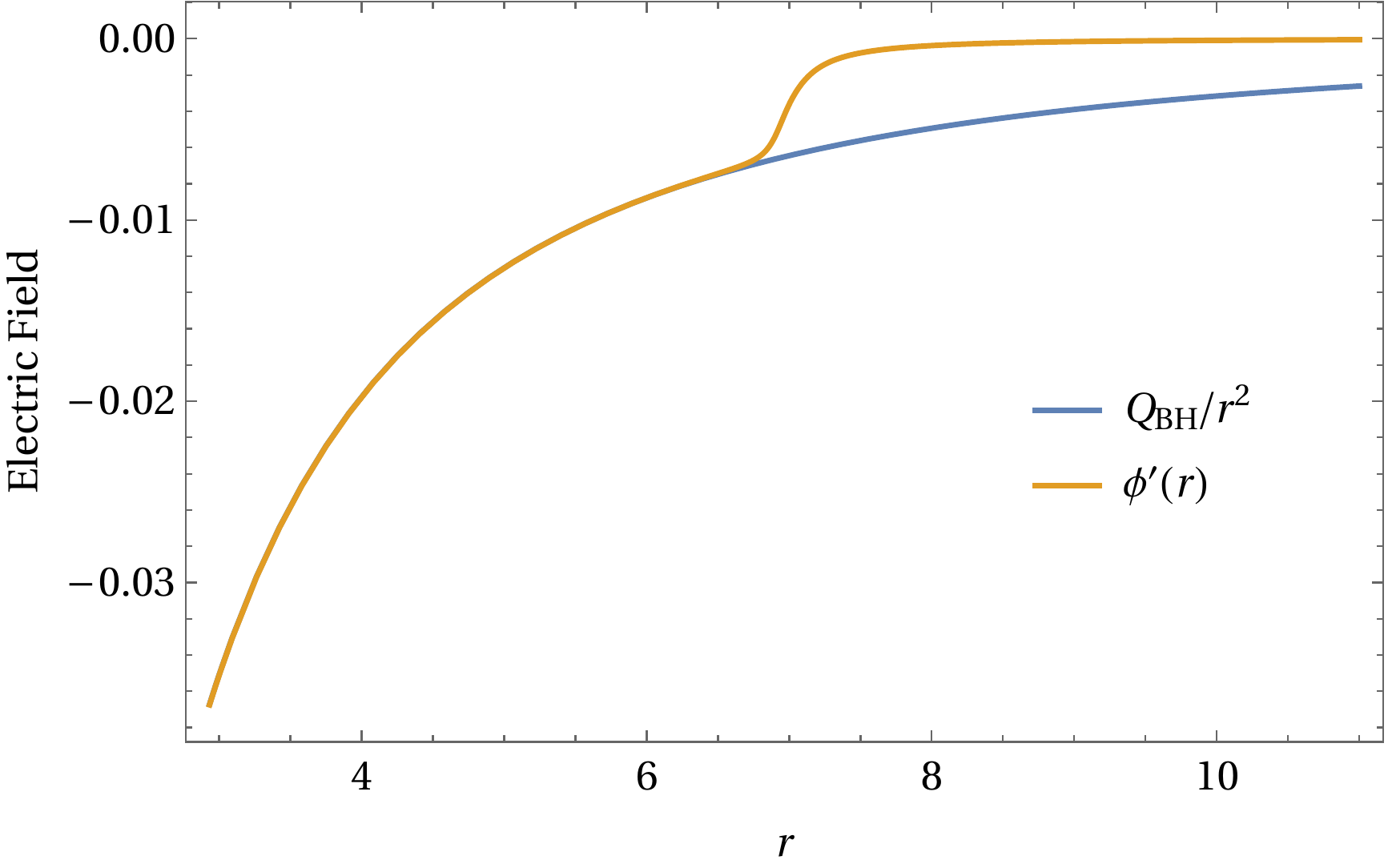}
    \qquad 
    \includegraphics[scale=0.45]{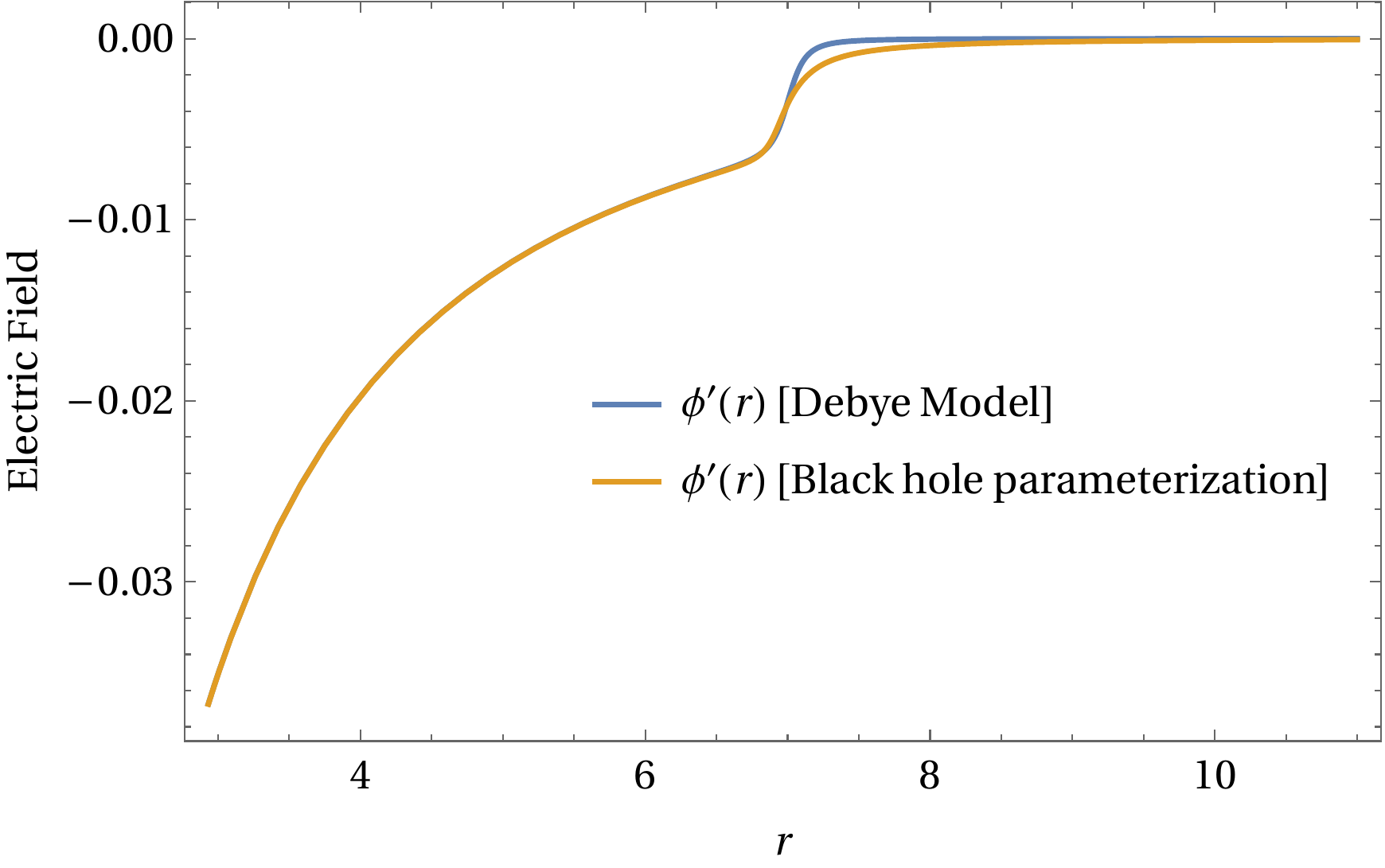}
    \caption{Electric field solution obtained with the BH
    parametrization for exaggerated parameter choices. Here, we choose
    $\tilde{Q}_{BH}=0.4081$, $M_{BH}=1$, $M_{ADM}=2$, $r_0=7$,
    $\lambda_0=1/8$, and $\delta r_{*0}=-0.1$. The parameters have been
    chosen so that $M_{BH}$, $Q_{BH}$ and $M_{ADM}$ match that of the
    Debye model solution in Fig. \ref{fig:ShieldingSolutionPhiE}. The
    parameter $\delta r_{*0}$ has been set so that the charge density is
    closer to the BH, as one might expect. The solution for the
    electric field is on the left, and a comparison with the Debye model
    solution is provided on the right.}
    \label{fig:BparEfieldSoln}
\end{figure*}
%%%%%%%%%%%%%%%%%%%%%%%%
%%%%%%%%%%%%%%%%%%%%%%%%

%%%%%%%%%%%%%%%%%%%%%%%%
%%%%%%%%%%%%%%%%%%%%%%%%
\begin{figure*}
    \centering
    \includegraphics[scale=0.45]{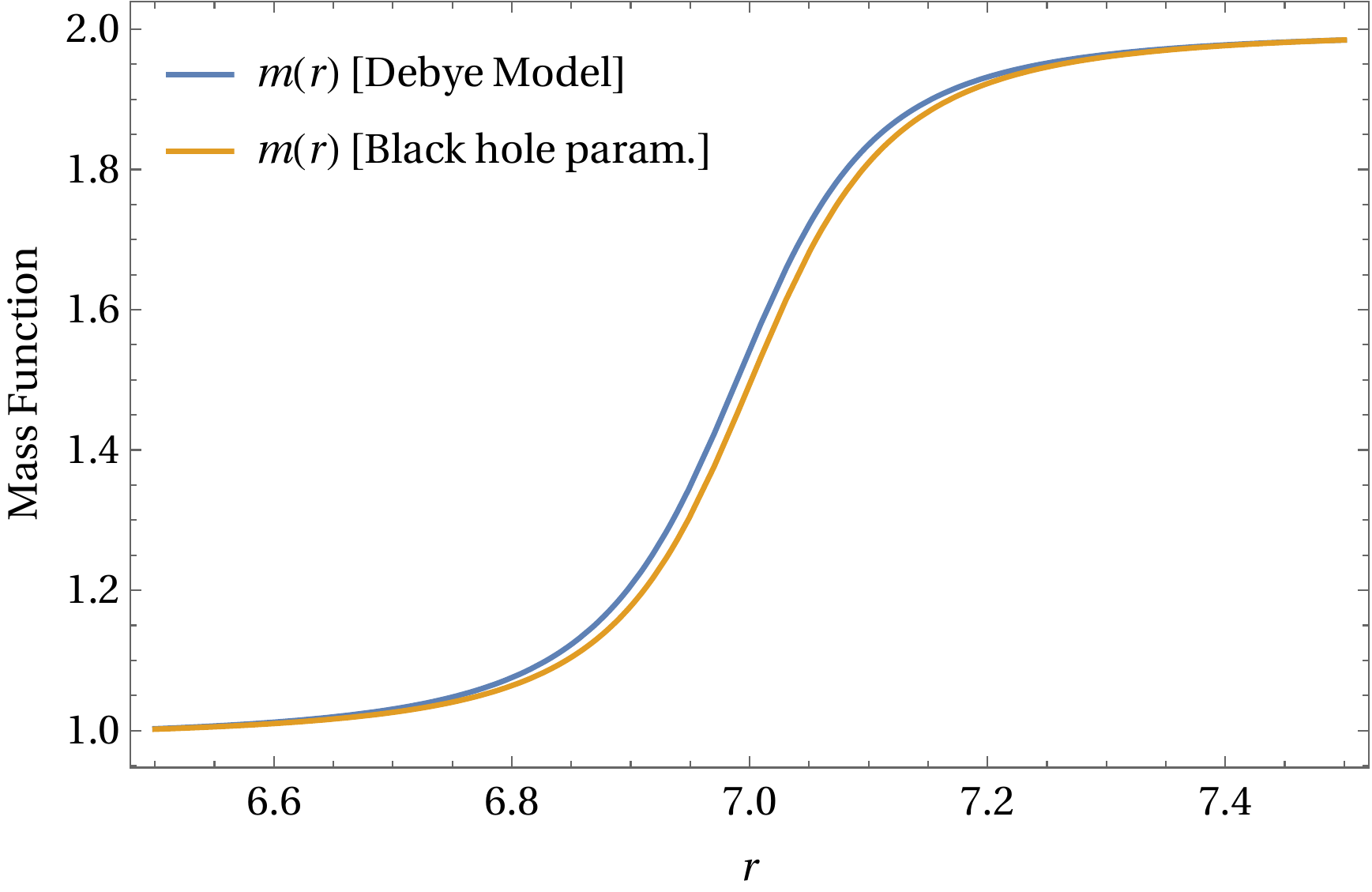}
    \qquad
    \includegraphics[scale=0.455]{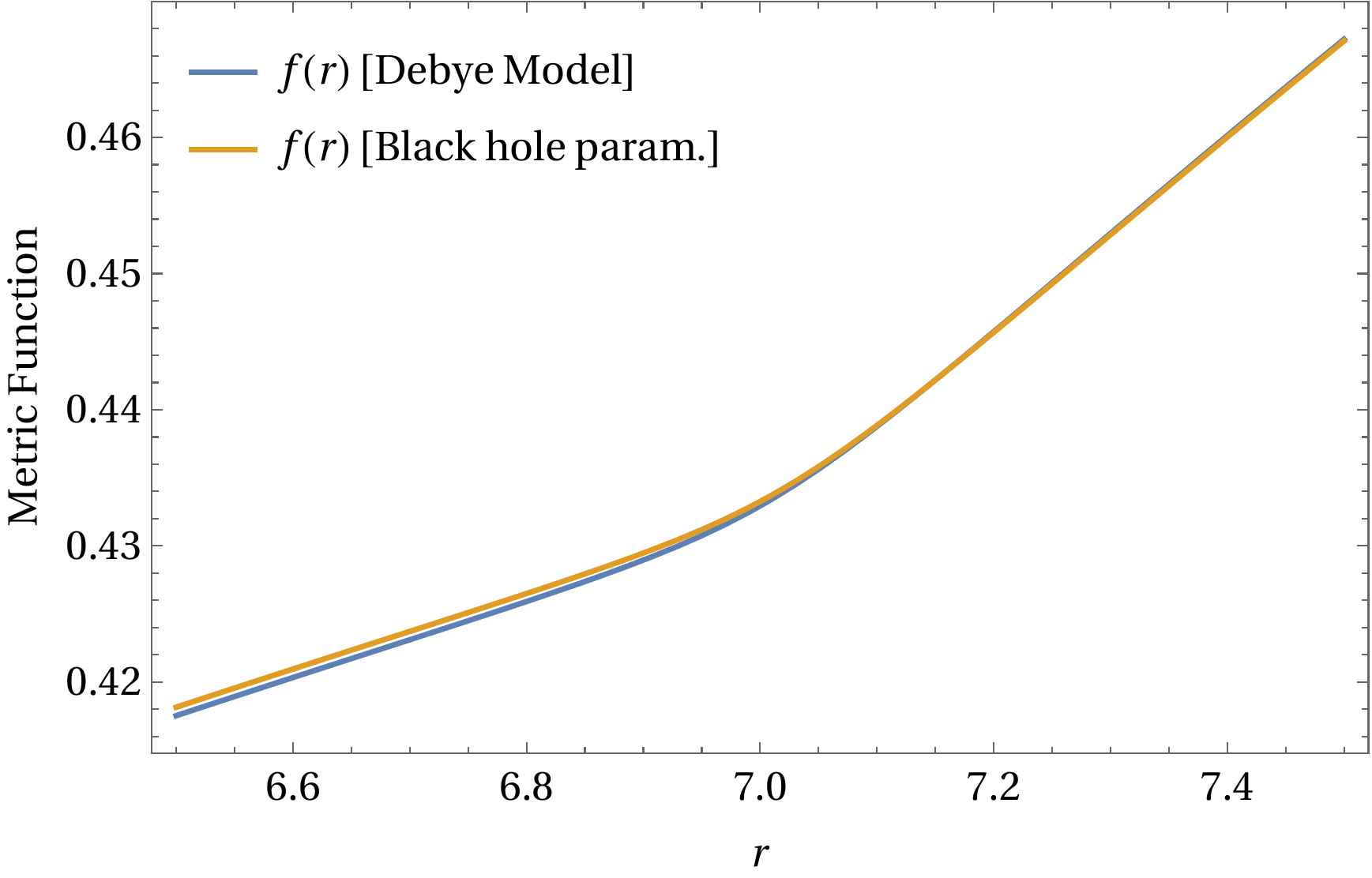}
    \caption{Comparison of geometries between the Debye model and BH
    parametrization solutions for exaggerated parameter choices in
    Figs. \ref{fig:ShieldingSolutionPhiE} and \ref{fig:BparEfieldSoln}.
    The mass function $m(r)$ is on the left, and the metric function
    $f(r)$ is on the right. The differences in the metric function
    $f(r)$ for $r<r_0=7$ can be attributed to the fact that the value
    for $Y_0=Y(r_H)=-0.5112$ here differs slightly from the value of
    $Y_0$ for the Debye model solution.}
    \label{fig:Bparmfcomp}
\end{figure*}
%%%%%%%%%%%%%%%%%%%%%%%%
%%%%%%%%%%%%%%%%%%%%%%%%

%%%%%%%%%%%%%%%%%%%%%%%%
%%%%%%%%%%%%%%%%%%%%%%%%
\begin{figure*}
    \centering
    \includegraphics[scale=0.45]{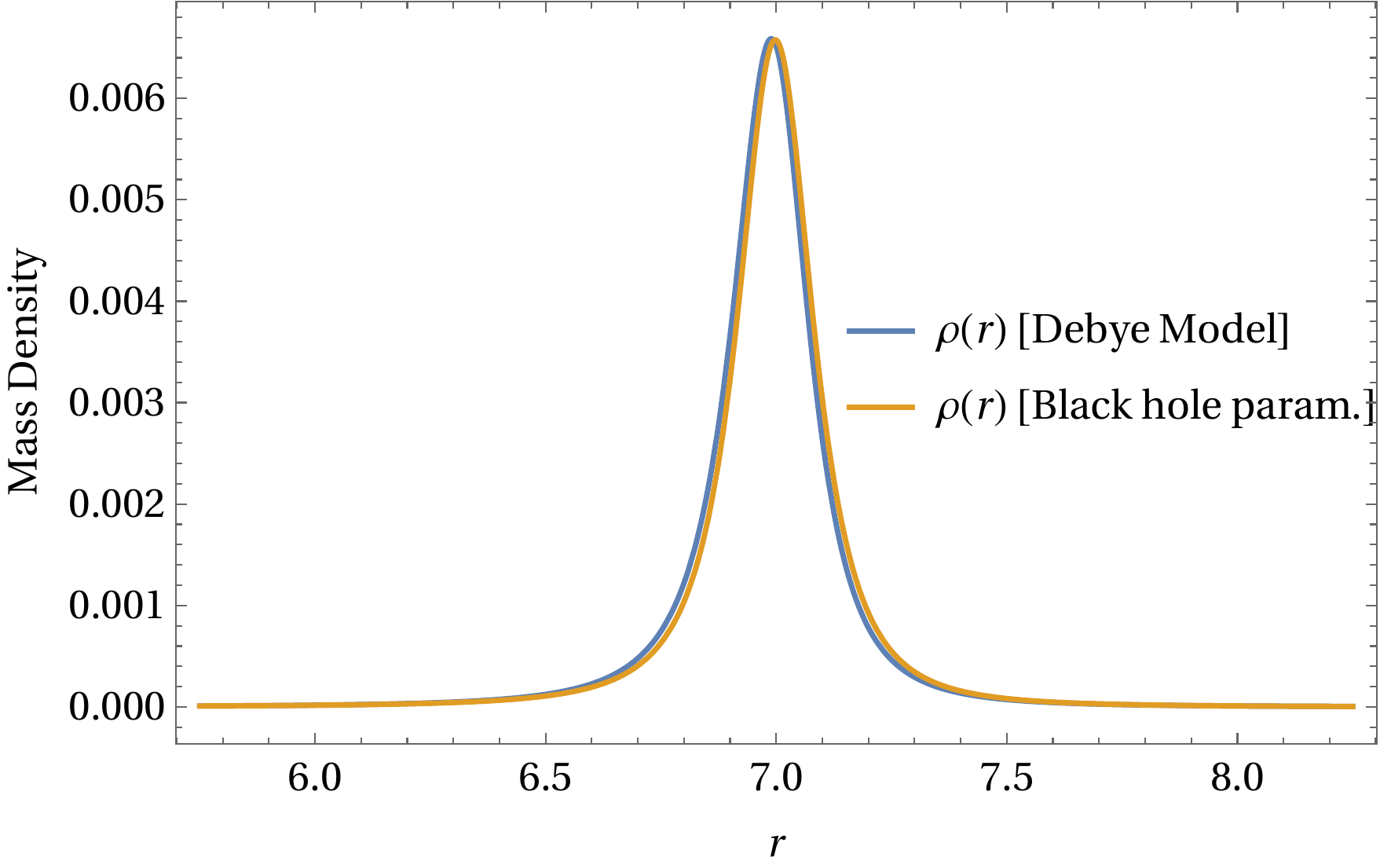}
    \qquad
    \includegraphics[scale=0.46]{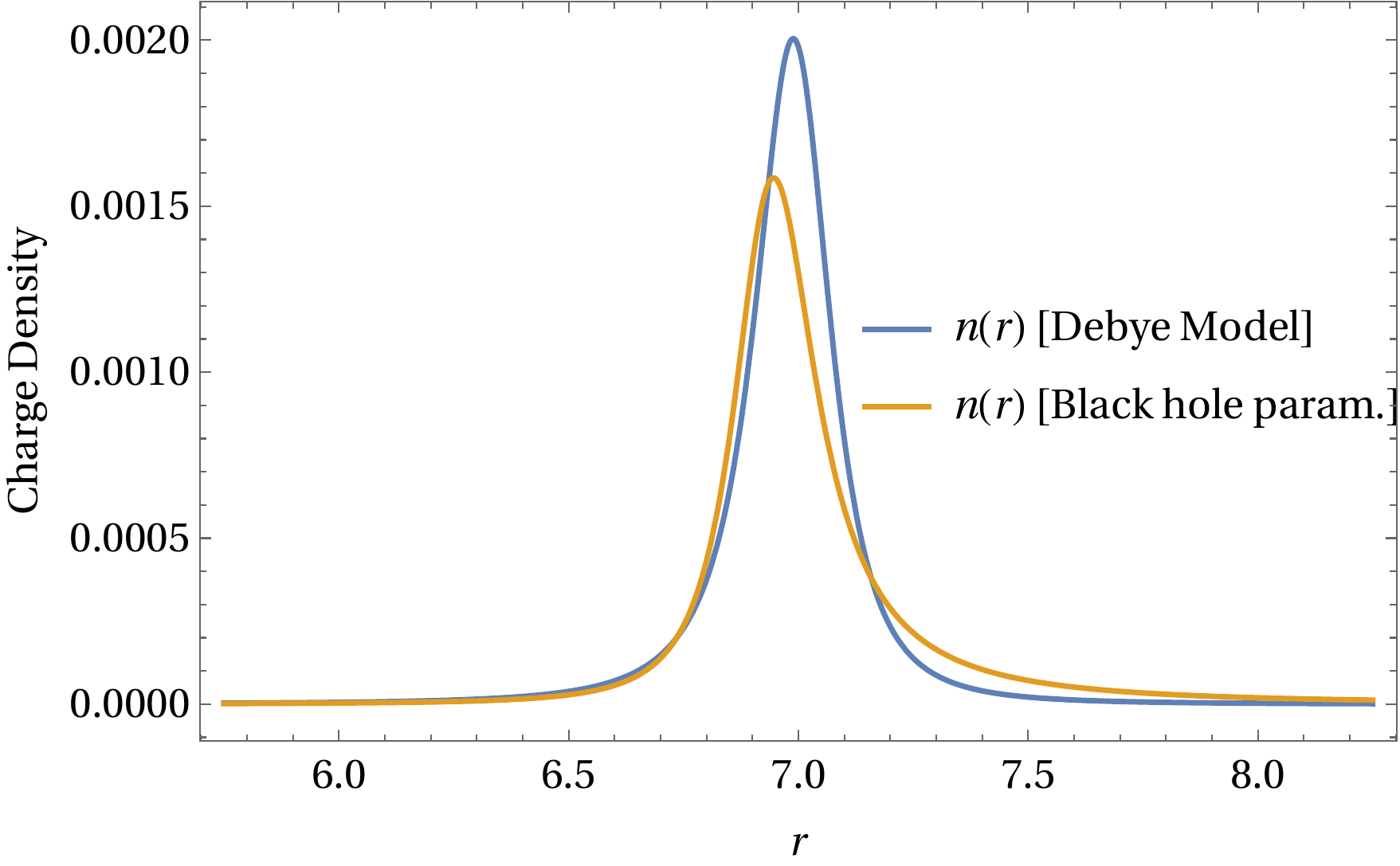}
    \caption{Comparison of density profiles between the Debye model and
    BH parametrization solutions for exaggerated parameter
    choices in Figs. \ref{fig:ShieldingSolutionPhiE} and
    \ref{fig:BparEfieldSoln}. The energy density $\rho(r)$ is on the left,
    and the charge density $n(r)$ is on the right. We recall the choice
    $\delta r_{*0}=-0.1$, which shifts the peak of the charge density
    profile $n(r)$ to a slightly lower radius; we do this because if
    $\delta r_{*0}$ is chosen to vanish, the BH parametrization
    tends to shift the peak of $n(r)$ to larger values of $r$ compared
    to the Debye model. The comparison plot for the tangential pressure
    profile is qualitatively similar to that of $n(r)$, so we omit it
    here.}
    \label{fig:Bpardenscomp}
\end{figure*}
%%%%%%%%%%%%%%%%%%%%%%%%
%%%%%%%%%%%%%%%%%%%%%%%%

One difficulty with the approach in the preceding section is that in
assuming a relationship between the density $n(r)$ and the potential
$\phi(r)$, the charge density $n(r)$ is formally sensitive to shifts in
the potential $\phi(r)$, though as we have argued, one can in principle
absorb such shifts into the parameter $\nu_0$ for a
Maxwell-Boltzmann-like distribution. In this section, we consider an
alternative approach which does not assume a particular relationship
between the density $n(r)$ and the potential $\phi(r)$. Instead we
perform a physically motivated reparametrization, and choose explicit
forms for a pair of functions --- such an approach provides a direct
phenomenological model for a shielded BH. 

We begin by reparameterizing the geometry, in order to bring out the
physics of these solutions. First of all, we choose the mass function
$m(r)$, as follows:
%%%%%%%%%%%%%%%%%%%%%%%%%%%%%%%%%%%%%%%%%
\begin{equation}\label{Eq:Mbh}
m(r) = M_{\rm BH}-\frac{Q_{\rm BH}^2}{2 r}e^{-Y_{0}} 
+ \left[\mathcal{M}(r)+\mathcal{Q}(r)\right]e^{-2Y_{0}}f_{\rm EV}^2~,
\end{equation}
%%%%%%%%%%%%%%%%%%%%%%%%%%%%%%%%%%%%%%%%%
and the energy density of the fluid as
%%%%%%%%%%%%%%%%%%%%%%%%%%%%%%%%%%%%%%%%%
\begin{equation}\label{Eq:rhobh}
\begin{aligned}
\rho (r) = \, & \frac{2}{8\pi r^{2}}\dfrac{d}{dr}\left[e^{-2 Y_0} \mathcal{M}(r)f_{\rm EV}^{2}\right]\\
= \, &\frac{2 f_{\rm EV}}{ e^{2 Y_0} 8\pi  r^2}\biggl[\mathcal{M}'(r) f_{\rm EV}
+4 \mathcal{M}(r) \left(\frac{M_{\rm BH}e^{Y_{0}}}{r^2}-\frac{Q_{\rm BH}^2}{r^3}\right)
\biggr],
\end{aligned}
\end{equation}
%%%%%%%%%%%%%%%%%%%%%%%%%%%%%%%%%%%%%%%%%
where we have introduced the radial function $f_{\rm EV}$ from Eq.
\eqref{Eq:evacsol}. Thus, we impose that the external mass distribution
is such that the geometry contains a central charged
Reissner-Nordstr\"{o}m-like BH, with mass $M_{\rm BH}$ and charge $\pm
Q_{\rm BH}e^{-Y_{0}/2}$. In vacuum, $\mathcal{Q}(r)=\mathcal{M}(r)=0$
and one recovers identically the geometry of Reissner-Nordstr\"{o}m BH,
since in that case $Y_{0}$ vanishes as well. 

Therefore, the $g^{rr}$ component of the metric reads
%%%%%%%%%%%%%%%%%%%%%%%%%%%%%%%%%%%%%%%%%
\begin{equation}\label{grr}
g^{rr}=\frac{f_{\rm EV}}{e^{Y_{0}}}\left[1-\frac{2\mathcal{M}(r)+2\mathcal{Q}(r)}{e^{Y_{0}} r}f_{\rm EV}\right]~.
\end{equation}
%%%%%%%%%%%%%%%%%%%%%%%%%%%%%%%%%%%%%%%%%
Finally, from Eq. \eqref{Eq:YDef}, it follows that the function $f(r)$,
i.e., the $g_{tt}$ component of the metric, becomes
%%%%%%%%%%%%%%%%%%%%%%%%%%%%%%%%%%%%%%%%%
\begin{equation}\label{Eq:fr}
\begin{aligned}
f(r) = \, & 
        \frac{f_{\rm EV} e^{Y(r)}}{e^{Y_{0}}}
        \left[1-\frac{2\mathcal{M}(r)+2\mathcal{Q}(r)}{e^{Y_{0}} r}f_{\rm EV}\right] ,
\end{aligned}
\end{equation}
%%%%%%%%%%%%%%%%%%%%%%%%%%%%%%%%%%%%%%%%%
where, the function $Y(r)$ satisfies the following first order
differential equation
%%%%%%%%%%%%%%%%%%%%%%%%%%%%%%%%%%%%%%%%%
\begin{equation}\label{Eq:Yp}
\begin{aligned}
\frac{dY}{dr}&=\frac{8\pi r \rho(r)}{g^{rr}}=\frac{2\dfrac{d}{dr}\left[\mathcal{M}(r)f_{\rm EV}^{2}\right]}{f_{\rm EV}\Big[e^{Y_0}r-\left\{2\mathcal{M}(r)+2\mathcal{Q}(r)\right\}f_{\rm EV}\Big]}
\\
&=
\frac{2 r^3 f_{\rm EV} \mathcal{M}'(r)-8 \mathcal{M}(r) \left(Q_{\rm BH}^2-e^{Y_0} r M_{\rm BH}\right)}
{e^{Y_0} r^4-2 (\mathcal{M}(r)+\mathcal{Q}(r)) r^3 f_{\rm EV} }
\end{aligned}
\end{equation}
%%%%%%%%%%%%%%%%%%%%%%%%%%%%%%%%%%%%%%%%%
Note that $(dY/dr)$ and hence $Y(r)$ are both finite in the limit,
$r\rightarrow r_{\pm}$, where $r_{\pm}$ are the solutions of the
algebraic equation, $f_{\rm RN}=0$. In particular, note that the
surfaces $r=r_{\pm}$ are null, since $g^{rr}$ vanishes there and so does
$(\partial/\partial t)^{\mu}(\partial/\partial t)_{\mu}$. Therefore, the
spacetime has an event horizon at $r=r_{+}=M_{\rm BH}+\sqrt{M_{\rm
BH}^{2}-Q_{\rm BH}^{2}e^{-Y_{0}}}$ and a Cauchy horizon at
$r=r_{-}=M_{\rm BH}-\sqrt{M_{\rm BH}^{2}-Q_{\rm BH}^{2}e^{-Y_{0}}}$.
These are precisely where the horizons of the Reissner-Nordstr\"{o}m BHs
are located, in this ``areal'' coordinate. Moreover, as evident from
Eq.~\eqref{Eq:rhobh}, on the event horizon the energy density $\rho$
identically vanishes, which is also a desirable property. Therefore, we
have traded the mass function $m(r)$ and the energy density $\rho(r)$ of
the charged cloud in terms of the functions $\mathcal{M}(r)$ and
$\mathcal{Q}(r)$. Note that, $Y(r)$ gets determined in terms of these
two functions and hence the metric element $g_{tt}$. Therefore, the
three unknowns $\{Y(r),m(r),\rho(r)\}$ reduce to the following two
unknowns $\{\mathcal{M}(r),\mathcal{Q}(r)\}$. 

On a different note, notice that $Y(r)$ satisfies a first order
differential equation whose integration needs to be done with the
boundary condition that $Y(r\rightarrow \infty)=Y_{\infty}$, which can
be chosen to be zero. However, due to the monotonicity of $Y'(r)$, it
follows that $Y(r_{+})<Y_{\infty}$ and hence $f(r)\neq g^{rr}$ on the
horizon, but will differ by an overall factor $\sim
e^{Y_{+}-Y_{\infty}}$. 

Let us now concentrate on the equations governing the electric
potential. First of all, the field equation presented in Eq.
\eqref{Eq:XEFE2} needs to be consistent with Eq. \eqref{Eq:Mbh} and Eq.
\eqref{Eq:rhobh}, which fixes the function $\psi(r)$ to be
%%%%%%%%%%%%%%%%%%%%%%%%%%%%%%%%%%%%%%%%%
\begin{equation}
\begin{aligned}
\psi(r)&=\frac{2m'-8\pi r^{2}\rho}{8\pi r^{2}}
\\
&=\frac{Q_{\rm BH}^{2}}{8\pi e^{Y_{0}} r^{4}}+\frac{2}{8\pi e^{2Y_{0}} r^{2}}
\frac{d}{dr}\left[\mathcal{Q}(r)f_{\rm EV}^{2} \right]~.
\end{aligned}
\end{equation}
%%%%%%%%%%%%%%%%%%%%%%%%%%%%%%%%%%%%%%%%%
Using Eq. \eqref{Eq:PsiDef}, we can express the electric potential as
%%%%%%%%%%%%%%%%%%%%%%%%%%%%%%%%%%%%%%%%%
\begin{equation}\label{Eq:hipsquare}
\phi'(r)^{2}=\frac{e^{Y(r)}}{e^{2Y_{0}}}\left[\frac{Q_{\rm BH}^{2}e^{Y_{0}}}{r^{4}}+\frac{2}{r^{2}}
\frac{d}{dr}\left[\mathcal{Q}(r)f_{\rm EV}^{2} \right]\right]~.
\end{equation}
%%%%%%%%%%%%%%%%%%%%%%%%%%%%%%%%%%%%%%%%%
We should point out that, as expected, the electric potential depends on
the charge of the BH, $Q_{\rm BH}$, and the energy density from the
electric field energy of the charged cloud, denoted by $\mathcal{Q}(r)$.
In absence of both the electric potential would vanish, as expected.

Finally, from Eq. \eqref{Eq:ddphi} the differential equation for the
electric potential $\phi(r)$ reads
%%%%%%%%%%%%%%%%%%%%%%%%%%%%%%%%%%%%%%%%%
\begin{equation}
\begin{aligned}
\phi''+\frac{2\phi'}{r}&=
\frac{\phi'}{r e^{2Y_0} g^{rr}}\dfrac{d}{dr}\left[\mathcal{M}(r)f_{\rm EV}^{2}\right]
+\frac{4\pi q\sqrt{f} n(r)}{g^{rr}}
\\
&=
\frac{\phi'}{r e^{2Y_0} g^{rr}}\dfrac{d}{dr}\left[\mathcal{M}(r)f_{\rm EV}^{2}\right]
+\frac{4\pi qn(r)}{\sqrt{g^{rr}}}e^{Y(r)/2}~.
\end{aligned}
\end{equation}
%%%%%%%%%%%%%%%%%%%%%%%%%%%%%%%%%%%%%%%%%
Using Eq. \eqref{Eq:Yp}, we can rewrite the above equation as
%%%%%%%%%%%%%%%%%%%%%%%%%%%%%%%%%%%%%%%%%
\begin{equation}\label{Eq:phipp}
\begin{aligned}
\phi''+\frac{2\phi'}{r}&=\frac{\phi' Y'}{2}
+\frac{4\pi qn(r)}{\sqrt{g^{rr}}}e^{Y(r)/2}~.
\end{aligned}
\end{equation}
%%%%%%%%%%%%%%%%%%%%%%%%%%%%%%%%%%%%%%%%%

With the reparametrization in hand, one can solve the system of
equations by specifying the mass distribution $\mathcal{M}(r)$ and the
charge distribution $\mathcal{Q}(r)$, based on their expected forms in
regions where the density $\rho(r)$ vanishes. Here, one assumes that the
density profile $\rho(r)$ is centered at a radius $r_0>r_{+}$, and has a
width of $\sigma_0$ [so that $\rho(r)$ vanishes for $|r-r_0|>\sigma_0$].
Near the horizon, one requires:
%%%%%%%%%%%%%%%%%%%%%%%%%%%%%%%%%%%%%%%%%
\begin{equation}\label{Eq:MfuncHorizon}
\begin{aligned}
\mathcal{M}(r\rightarrow r_{+})&=0~,
\\
\mathcal{Q}(r\rightarrow r_{+})&=0~.
\end{aligned}
\end{equation}
%%%%%%%%%%%%%%%%%%%%%%%%%%%%%%%%%%%%%%%%%
At radii $r \gg r_0+\sigma_0$ one has $\mathcal{M}(r \gg r_0+\sigma_0
)\approx\mathcal{M}_L(r)$, where
%%%%%%%%%%%%%%%%%%%%%%%%%%%%%%%%%%%%%%%%%
\begin{equation}\label{Eq:MQfuncLargeRad}
\begin{aligned}
\mathcal{M}_L(r)
&:=
\frac{r^4 \left(M_{\rm ADM}-M_{\rm BH}-\mathcal{Q}_{\infty }\right)}{\left(r \left(r-2 M_{\rm BH}\right)+\tilde{Q}_{\rm BH}^2\right)^2}~,
\end{aligned}
\end{equation}
%%%%%%%%%%%%%%%%%%%%%%%%%%%%%%%%%%%%%%%%%
which can be obtained by solving Eq. \eqref{Eq:Mfunction} when
$Y(r)=Y'(r)=0$; we define an effective charge $\tilde{Q}_{\rm BH}^2:=Q_{\rm BH}^2 e^{-Y_0}$,
and the quantity $\mathcal{Q}_{\infty }$ is an integration constant
which coincides with the value of $\mathcal{Q}(r)$ at large $r$. The
latter can be inferred by solving Eq. \eqref{Eq:hipsquare} for
$\mathcal{Q}(r)$ [assuming $\phi'(r \gg r_0+\sigma_0)=0$] and solving
Eq. \eqref{Eq:ZMfunction} for $\mathcal{M}(r)$. One may then write:
%%%%%%%%%%%%%%%%%%%%%%%%%%%%%%%%%%%%%%%%%
\begin{equation}\label{Eq:Mfunc}
\begin{aligned}
\mathcal{M}(r)=&\sigma_M\left(\frac{r_{*}-r_{*0}}{\sigma_{*0}}\right) \mathcal{M}_L(r)~,
\end{aligned}
\end{equation}
%%%%%%%%%%%%%%%%%%%%%%%%%%%%%%%%%%%%%%%%%
where $\sigma_M(x)$ is a sigmoidal function, with the property that
$\sigma_M(-\infty)=0$ and $\sigma_M(+\infty)=1$. Here $r_{*}$ is a
tortoise coordinate defined with respect to the horizon radius of
$f_{\rm EV}$, such that $r_{*}(r)=r+r_H \ln(r/r_H-1)$ and
$r_{*0}=r_{*}(r_0)$. We define $\sigma_{*0}:=\sigma_{0} r'_{*}(r_0)$,
assuming $\sigma_{0} \ll r_0$.
%, such that $(dr/dr_{*})=f_{\rm EV}$. 
Given a solution to Eq. \eqref{Eq:hipsquare} for $r \gg r_0+\sigma_0$,
one can assume a similar expression for $\mathcal{Q}(r)$, but it is
perhaps more appropriate to specify the derivative of $\mathcal{Q}(r)$
in the following manner:
%%%%%%%%%%%%%%%%%%%%%%%%%%%%%%%%%%%%%%%%%
\begin{equation}\label{Eq:hipsquare2}
e^{-2Y_{0}}\frac{d}{dr}\left[\mathcal{Q}(r)f_{\rm EV}^{2} \right] = - \sigma_Q\left(\frac{r_{*}-r_{*0}-\delta r_{*0}}{\sigma_{*0}}\right) \frac{\tilde{Q}_{\rm BH}^{2}}{2r^{2}},
\end{equation}
%%%%%%%%%%%%%%%%%%%%%%%%%%%%%%%%%%%%%%%%%
so that, upon comparison with Eq. \eqref{Eq:hipsquare} and setting
$\delta r_{*0}=0$, one has $|\phi'(r < r_0-\sigma_0)|\approx
|Q_{BH}|/r^2$ and $\phi'(r > r_0+\sigma_0)\approx 0$. The adjustable
parameter $\delta r_{*0}$ has been introduced so that one has control
over the position of the charge distribution.

Given the mass distribution $\mathcal{M}(r)$ and the charge distribution
$\mathcal{Q}(r)$, one can then solve for $Y(r)$ from Eq. \eqref{Eq:Yp},
then $\phi'(r)$ from Eq. \eqref{Eq:hipsquare}, the number density $n(r)$
from Eq. \eqref{Eq:phipp} and finally the tangential pressure $\bar{P}$
from Eq. \eqref{Eq:XFl3}. Plots for $\phi'(r)$ and the metric functions
$m(r)$ and $f(r)$ are provided in the respective Figs.
\ref{fig:BparEfieldSoln} and \ref{fig:Bparmfcomp}, and the density
profile comparisons are provided in Fig. \ref{fig:Bpardenscomp}.

\begin{figure*}
    \centering
    \includegraphics[scale=0.45]{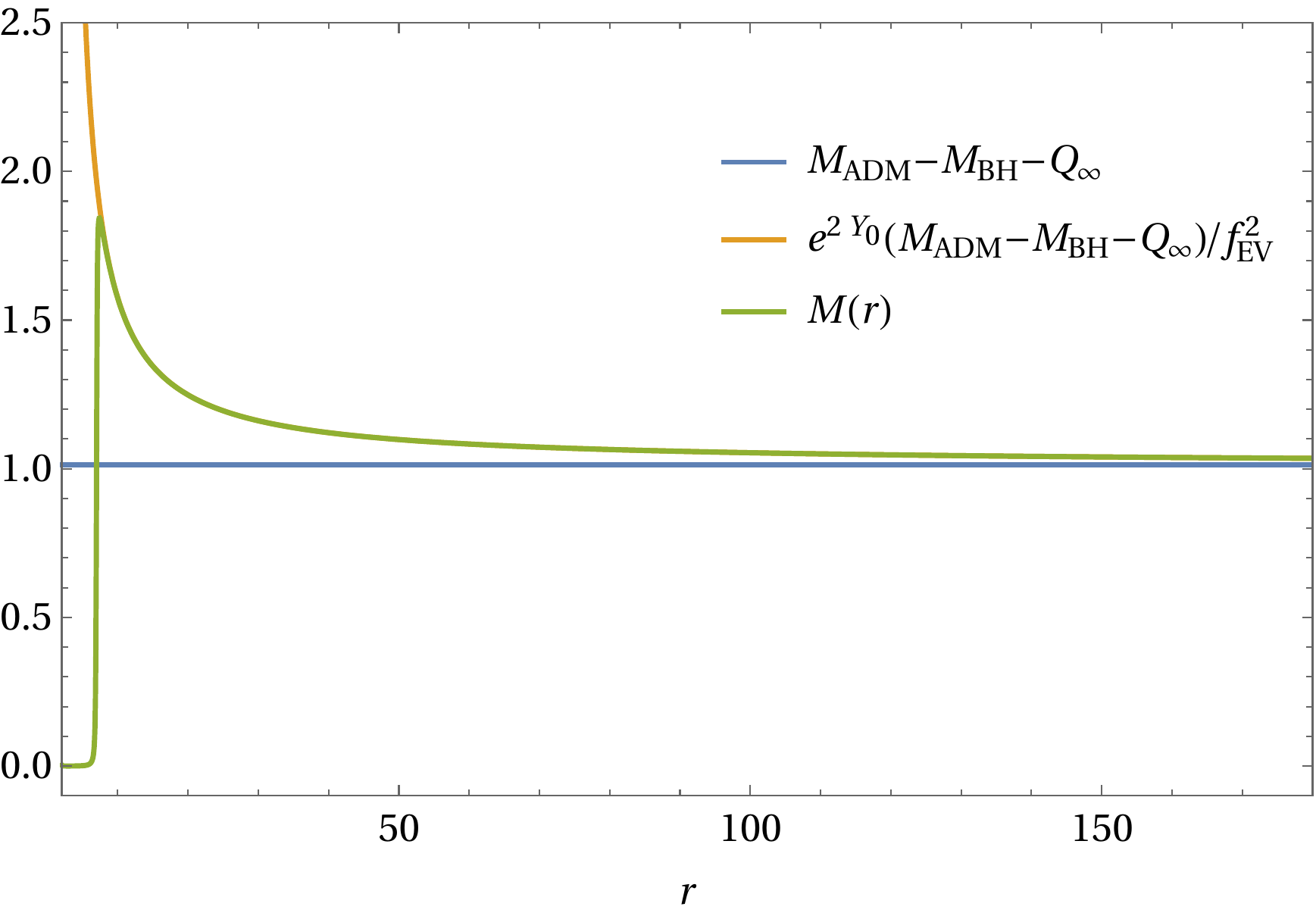}
    \qquad
    \includegraphics[scale=0.47]{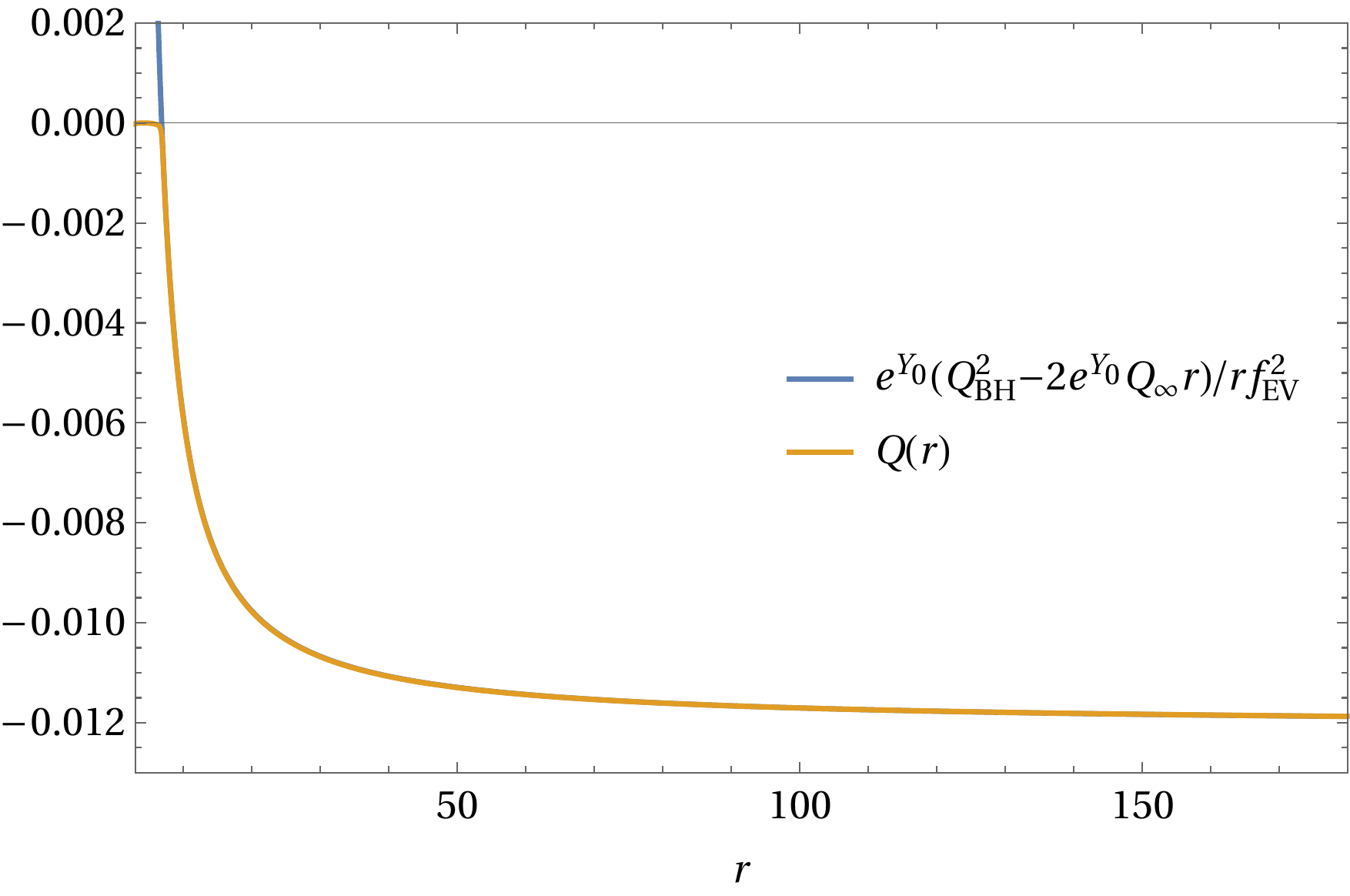}
    \caption{Plots of the functions $\mathcal{M}(r)$ and
    $\mathcal{Q}(r)$ obtained from the Debye model solutions in Figs.
    \ref{fig:ShieldingSolutionPhiE} and \ref{fig:ShieldingSolutionmfE}.
    The profile for $\mathcal{M}(r)$ is the on the left, and the profile
    for $\mathcal{Q}(r)$ is on the right. The corresponding profiles for
    the BH parametrization are not shown here, as they are
    nearly indistinguishable from the plots shown, and are qualitatively
    similar near $r=r_0$.} 
    \label{fig:DEMQ}
\end{figure*}

To compare the results obtained with the present parametrization with
that obtained by the method in Sec. \ref{Sec:Debye}, we note that the
following equations hold true:
%%%%%%%%%%%%%%%%%%%%%%%%%%%%%%%%%%%%%%%%%
\begin{equation}\label{Eq:ZMfunction}
\begin{aligned}
\left[\mathcal{M}(r)+\mathcal{Q}(r)\right] \frac{f_{\rm EV}^2}{e^{2Y_{0}}}=m(r) + \frac{Q_{\rm BH}^2}{2 e^{Y_{0}} r} -  M_{\rm BH}~,
\end{aligned}
\end{equation}
\begin{equation}\label{Eq:Mfunction}
\begin{aligned}
\dfrac{d}{dr}\left[e^{-2Y_{0}}f_{\rm EV}^2 \, \mathcal{M}(r)\right]=\frac{r}{2}f(r)e^{Y(r)} Y'(r)~,
\end{aligned}
\end{equation}
%%%%%%%%%%%%%%%%%%%%%%%%%%%%%%%%%%%%%%%%%
which one may solve for $\mathcal{M}(r)$ and $\mathcal{Q}(r)$, given
$m(r)$, $f(r)$, and $Y(r)$. Specifically, one solves Eq.
\eqref{Eq:Mfunction} for the mass parameter $\mathcal{M}(r)$ and then
Eq. \eqref{Eq:ZMfunction} provides the charge function $\mathcal{Q}(r)$.
Alternatively, if one has $\phi(r)$ and $Y(r)$ in hand, one can first
solve Eq. \eqref{Eq:hipsquare} to obtain $\mathcal{Q}(r)$, then Eq.
\eqref{Eq:ZMfunction} provides the mass function $\mathcal{M}(r)$. We
illustrate in Fig. \ref{fig:DEMQ} profiles for $\mathcal{M}(r)$ and
$\mathcal{Q}(r)$ obtained from the Debye model solution of Fig.
\ref{fig:ShieldingSolutionPhiE}.

%%%%%%%%%%%%%%%%%%%%%%%%
%%%%%%%%%%%%%%%%%%%%%%%%
%%%%%%%%%%%%%%%%%%%%%%%%
%%%%%%%%%%%%%%%%%%%%%%%%

%-----------------------------------------------------------------------
%-----------------------------------
%-----------------
%--------
%---
%-
%
%
%-
%---
%--------
%-----------------
%-----------------------------------
%-----------------------------------------------------------------------

%=======================================================================
\section{Physical Properties of the spacetime} \label{Sec:PhysProps}
%=======================================================================
Having derived the metric elements of a charged BH in a charged cloud
distribution, within a static and spherically symmetric context, in a
closed form, let us note down a few of its interesting physical
properties. These will include the location of the photon sphere
\cite{Claudel:2000yi}, the location of the innermost stable circular
orbits, the angular frequency and the Lyapunov exponents associated with
the photon sphere. As we have already noted there will be an event
horizon, located at $r_{+}=M_{\rm BH}+\sqrt{M_{\rm BH}^{2}-Q_{\rm
BH}^{2}e^{-Y_{0}}}$, on which the energy density $\rho$ identically
vanishes, while the tangential pressure $\bar{P}$ remains finite. 

The photon sphere, or, the unstable circular photon orbits are given by
the condition, $rf'=2f$, which from Eq. \eqref{Eq:XEFE1} and Eq.
\eqref{Eq:YDef} yields
%%%%%%%%%%%%%%%%%%%%%%%%%%%%%%%%%%%%%%%%%
\begin{equation}
\begin{aligned}
\frac{8\pi r^{2}\rho}{r-2m(r)}&=\frac{rf'+f}{rf}-\frac{1-2m(r)'}{r-2m(r)}
\\
&=\frac{3}{r}-\frac{1-2m(r)'}{r-2m(r)}~.
\end{aligned}
\end{equation}
%%%%%%%%%%%%%%%%%%%%%%%%%%%%%%%%%%%%%%%%%
Using Eq. \eqref{Eq:XEFE2}, we can express the above equation as
%%%%%%%%%%%%%%%%%%%%%%%%%%%%%%%%%%%%%%%%%
\begin{equation}
\frac{8\pi r^{2}\rho}{r-2m(r)}=\frac{2}{r}\left(\frac{r-3m(r)}{r-2m(r)}\right)+\frac{8\pi r^{2}\rho +8\pi r^{2}\psi}{r-2m}~,
\end{equation}
%%%%%%%%%%%%%%%%%%%%%%%%%%%%%%%%%%%%%%%%%
yielding the following expression satisfied by the radius of the photon
sphere:
%%%%%%%%%%%%%%%%%%%%%%%%%%%%%%%%%%%%%%%%%
\begin{equation}\label{phsph}
\begin{aligned}
r&=3m(r)-\frac{8\pi r^{3}\psi}{2}
\\
&=3M_{\rm BH}-\frac{2Q_{\rm BH}^{2}}{e^{Y_{0}}r}+3\frac{f_{\rm EV}^{2}}{e^{2Y_{0}}}\left[\mathcal{M}(r)+\mathcal{Q}(r)\right]
\\
&\qquad \qquad - \frac{r}{e^{2Y_{0}}}\frac{d}{dr}\left[\mathcal{Q}(r)f_{\rm EV}^{2}\right]~.
\end{aligned}
\end{equation}
%%%%%%%%%%%%%%%%%%%%%%%%%%%%%%%%%%%%%%%%%
Note that, in the absence of any charge on the BH, as well as the
absence of charged cloud leads to the well known radius of the photon
sphere, at $r_{\rm ph}=3M_{\rm BH}$. However, presence of a charged
matter distribution and charge on the BH itself shifts the photon
circular orbit to a different value. In the case where
$\mathcal{M}(r)=\mathcal{Q}(r)=0$, one has a photon sphere radius:
%%%%%%%%%%%%%%%%%%%%%%%%%%%%%%%%%%%%%%%%%
\begin{equation}\label{phsphEV}
\begin{aligned}
r_{\rm ph, EV}&=\frac{1}{2} \left(3 M_{\rm BH}+\sqrt{9 M_{\rm BH}^2-8 e^{-Y_{0}} Q_{\rm BH}^2}\right),
\end{aligned}
\end{equation}
%%%%%%%%%%%%%%%%%%%%%%%%%%%%%%%%%%%%%%%%%
which differs from the usual Reissner-Nordstr\"{o}m result due to the
presence of the factor $e^{-Y_{0}}$.
%%%%%%%%%%%%%%%%%%%%%%%%%%%%%%%%%%%%%%%%%
Also the critical impact parameter $b_{\rm crit}\equiv
(L_{\rm crit}/E_{\rm crit})$, related to the capturing of null geodesics
by the BH with the charged cloud system, would correspond to $E_{\rm
crit}=V_{\rm max}$, where $V_{\rm max}$ corresponds to the maximum of
the effective potential. For photons, the maxima corresponds to the
photon sphere, and, hence, 
%%%%%%%%%%%%%%%%%%%%%%%%%%%%%%%%%%%%%%%%%
\begin{equation}
b_{\rm crit}=\frac{r_{\rm ph}}{\sqrt{f(r_{\rm ph})}}=\frac{r_{\rm ph}^{3/2}}{\sqrt{r_{\rm ph}-2m(r_{\rm ph})}}~e^{-Y(r_{\rm ph})/2}~,
\end{equation}
%%%%%%%%%%%%%%%%%%%%%%%%%%%%%%%%%%%%%%%%%
where $r_{\rm ph}$ corresponds to the location of the photon sphere,
which is a solution to Eq. \eqref{phsph}. The angular frequency and the
Lyapunov exponent associated with the photon sphere reads
%%%%%%%%%%%%%%%%%%%%%%%%%%%%%%%%%%%%%%%%%
\begin{align}
\Omega_{\rm ph}&=\frac{1}{b_{\rm crit}}~;
\\
\lambda_{\rm ph}&=\sqrt{\left(1-\frac{2m(r_{\rm ph})}{r_{\rm ph}}\right)\left(\frac{2f(r_{\rm ph})-r_{\rm ph}^{2}f''(r_{\rm ph})}{2r_{\rm ph}^{2}} \right)}~.
\end{align}
%%%%%%%%%%%%%%%%%%%%%%%%%%%%%%%%%%%%%%%%%
These provide the basic properties of the photon sphere. In particular,
the angular frequency $\Omega_{\rm ph}$ and Lyapunov exponent
$\lambda_{\rm ph}$ are respectively related to the real and imaginary
parts of the quasinormal mode frequencies in the eikonal limit ($n$ and
$l$ being nonnegative integers):
\begin{align}
\omega_{\rm ps}&=l \Omega_{\rm ph} - i (n+1/2) |\lambda_{\rm ph}|, \
\qquad l \gg 1.
\end{align}
%%%%%%%%%%%%%%%%%%%%%%%%%%%%%%%%%%%%%%%%%
Therefore, the above
expressions enable us to compute the quasinormal modes of the charged
cloud surrounding a charged BH system, in the large angular momentum
limit. So far, all the results we have derived are analytic in nature.
However, for obtaining a solution to the electrostatic potential, as
well as to the tangential pressure and the metric elements, we need
solutions for the differential equations presented in the previous
section. This in turn will enable us to derive the location of the
photon sphere from Eq. \eqref{phsph} and hence its related properties.

Values for the photon sphere radii $r_{\rm ph}$ and values for the
quantities characterizing null geodesics near $r_{\rm ph}$ for the
numerical solutions described in the preceding sections are given in
Table \ref{fig:PropNGPS}. We note that since the energy density becomes
sparse in the vicinity of the photon sphere (as we have chosen the
energy density to vanish below the ISCO), the photon sphere mode
frequencies for quasinormal modes in the Eikonal limit are equivalent to
those of an electrovacuum Reissner-Nordstr\"{o}m BH with a rescaled time
coordinate. However, since $e^{Y_0}\neq 1$ below the cloud, the
quasinormal mode frequencies as seen by an observer at infinity will
differ between that of a ``naked'' charged BH, and one that is
``clothed'' by a plasma.

\begin{table}
\begin{tabular}{
|p{2.8cm}||p{1.2cm}|p{1.2cm}|p{1.3cm}|p{1.3cm}|  }
 \hline
 \multicolumn{5}{|c|}{Properties of null geodesics near photon sphere}
 \\
 \hline
  & $r_{\rm ph}$ & $b_{\rm crit}$ & $\Omega_{\rm ph}$ & $\lambda_{\rm ph}$\\
 \hline
 Realistic values & $2.99994$ & $5.53852$ & $0.180554$ & $0.180552$ \\
 Exaggerated values & $2.88411$ & $6.51911$ &
 $0.153395$ & $0.150292$ \\
 \hline
 Realistic RN & $2.99995$ & $5.19608$ & $0.192453$ & $0.192451$ \\
 Exaggerated RN & $2.93153$ & $5.10730$ &
 $0.195798$ & $0.193511$ \\
 \hline
\end{tabular}
  \caption{The row termed ``Realistic values'' corresponds to the
  parameter choices of the Debye model solution in Fig.
  \ref{fig:ShieldingSolutionPhiR}. The row termed ``Exaggerated values''
  corresponds to the parameter choices of the Debye model solution in
  Fig. \ref{fig:ShieldingSolutionPhiE}. The rows termed ``Realistic RN''
  and `Exaggerated RN'' correspond to the same parameter choices for
  $M_{BH}$ and $Q_{BH}$ as the respective rows ``Realistic values'' and
  ``Exaggerated values'', except with the Reissner-Nordstr\"{o}m value for
  the factor $e^{Y_0} = 1$. Regardless of the model, the results are the
  same (up to precision errors) near the photon sphere radius $r_{\rm
  ph}$, as the energy density for the cases considered becomes sparse at
  the photon radius; essentially, $r_{\rm ph}$ is the photon radius
  obtained from $f_{\rm EV}$. }
\label{fig:PropNGPS}
\end{table}

%-----------------------------------------------------------------------
%-----------------------------------
%-----------------
%--------
%---
%-
%
%
%-
%---
%--------
%-----------------
%-----------------------------------
%-----------------------------------------------------------------------

%=======================================================================
\section{Conclusions}
%=======================================================================
The exact mechanism shielding a charged BH has remained elusive to
date. In this work we present a comprehensive understanding of the
screening mechanism of spherically symmetric charged BH surrounded by a
spherically symmetric charged matter distribution. Since there is no
radial outflow, as consistent with the galactic matter distribution, we
also consider an anisotropic matter distribution surrounding the BH
having only energy density and tangential stress, all functions of the
radial coordinate alone. The resulting system has---(a) two metric
degrees of freedom, the $g_{tt}$ and $g^{rr}$ components, (b) the energy
density and tangential pressure of the matter distribution, (c) the
electrostatic potential and (d) the charge density---in total six
functions of the radial coordinate alone. Since the energy and charge
densities are independent, our model can may be regarded as a rather
general effective description for charge overdensities in plasmas
surrounding charged BHs. However, the Einstein equations along with the
Maxwell equations provide four equations among these six
variables---this is not particularly surprising, as our model might be
regarded as an effective description for a multispecies fluid, in which
case one must supply equations of state for each fluid species. Thus in
order to close this system of equations, we need two more relations
among these variables. For that we make two possible choices---(a) we
provide a relation between the number density of the charged particles
with the electrostatics potential along with a suitable choice for the
ratio of the $g_{tt}$ and $g^{rr}$ components of the metric,
alternatively, (b) we fix the mass function $\mathcal{M}(r)$ and the
charge function $\mathcal{Q}(r)$ in the parametrized form of the
geometry, describing the charged matter, surrounding the charged BH.
Both of these approaches results in an electric field decaying
sufficiently rapidly; see, e.g., Fig. \ref{fig:BparEfieldSoln}. In
particular, fixing the relation between the number density of charged
particles with the electrostatic potential to be linear, results in a
much faster exponential decay, than a steep power law decay in the
parametrized case. Nonetheless, for generic choices of parameters, the
screening effect of the electric field remains, such that after moving a
small distance into the charged cloud, the electric field almost
diminishes to zero and hence it appears that overall the system is
uncharged, though there is a charged BH inside. 

This mechanism is intuitive, but has never been derived in an explicit
manner. Here, we directly solve the full nonlinear Einstein-Maxwell
system, by two independent methods as prescribed above: in doing so, we
obtain a description that accounts for the nonlinearities of general
relativity. In both of these cases, such a screening mechanism appears
generically, providing a direct proof that indeed the electric field of
a charged BH almost vanishes immediately after the internal surface of
the charged cloud is crossed. Since, in most of the astrophysical
scenario the BHs are surrounded by charged plasma, it follows that
astrophysically, i.e., from a large distance from the central BH, it
will always appear to be neutral. This is why astrophysical BHs appears
to be uncharged, due to the screening mechanism derived here. 

The matter distributions we have considered in this analysis are assumed
to vanish for radii below the innermost stable circular orbit for
uncharged particles. Consequently, the physical properties of null
geodesics near the photon sphere are essentially those of a charged BH
in electrovacuum, but we have found that the presence of the plasma
cloud changes the redshift factor, so that the eikonal limit quasinormal
modes seen by a faraway observer will differ from an ``unclothed''
charged BH. Still, we have presented general formulas for the angular
frequency and Lyapunov exponent in case one wishes to extend the
analysis to charged Einstein cluster models incorporating unstable
circular orbits near the photon sphere.

There are several future applications of the results discussed above.
The most immediate one would be generalization to Kerr-Newman BHs and
show that the screening mechanism continues to hold in the presence of
rotation as well. This is important, since all the astrophysical BHs are
supposed to have nonzero rotation. It would be interesting to observe
effects of the BH charge on the gravitational waves emanating from
perturbation of the same as it propagates through the surrounding
charged matter. In particular, whether the screening effect will also
affect the gravitational wave needs to be explored. It would also be
worthwhile to explore if similar results hold true for charges induced
by gravitational interactions as well, e.g., what happens to scalar
hairs, or hairs inherited from higher spacetime dimensions. These issues
will be addressed elsewhere. 

%=======================================================================

%-----------------------------------------------------------------------
%-----------------------------------
%-----------------
%--------
%---
%-
%
%
%-
%---
%--------
%-----------------
%-----------------------------------
%-----------------------------------------------------------------------

%=======================================================================

\begin{acknowledgments}
J.C.F thanks the Niels Bohr International Academy for hosting a visit
during which part of this research was performed, and acknowledges
financial support from FCT---Funda\c c\~ao para a Ci\^encia e a
Tecnologia of Portugal Project No.~UIDB/00099/2020.
V.C.\ is a Villum Investigator and a DNRF Chair, supported by VILLUM
FONDEN (grant no.~37766) and by the Danish Research Foundation. V.C.\
acknowledges financial support provided under the European Union's H2020
ERC Advanced Grant ``Black holes: gravitational engines of discovery''
grant agreement no.\ Gravitas–101052587.
This project has received funding from the European Union's Horizon 2020
research and innovation programme under the Marie Sklodowska-Curie grant
agreement No 101007855. We acknowledge financial support provided by
FCT/Portugal through grants 2022.01324.PTDC, PTDC/FIS-AST/7002/2020,
UIDB/00099/2020 and UIDB/04459/2020. 
Research of S.C. is funded by the INSPIRE Faculty fellowship from DST,
Government of India (Reg. No. DST/INSPIRE/04/2018/000893) and by the
Start-Up Research Grant from SERB, DST, Government of India (Reg. No.
SRG/2020/000409).
\end{acknowledgments}

%=======================================================================

%=======================================================================
%		BIBLIOGRAPHY
%=======================================================================

%-----------------------------------------------------------------------
%-----------------------------------
%-----------------
%--------
%---
%-

\bibliography{ref}

%-
%---
%--------
%-----------------
%-----------------------------------
%-----------------------------------------------------------------------

\end{document}